\documentstyle[12pt]{article}
\newcommand{\be}{\begin{equation}}
\newcommand{\ee}{\end{equation}}
\newcommand{\bea}{\begin{eqnarray}}
\newcommand{\eea}{\end{eqnarray}}
\newcommand{\nn}{\nonumber}
\newcommand{\D}{\displaystyle}

\newcommand{\ha}{\hat{a}}

\newcommand{\had}{\hat{a}^\dagger}

\newcommand{\hV}{\hat{V}}

\newcommand{\bb}{{\bf b}}
\newcommand{\ket}[1]{{|#1\rangle}}
\newcommand{\bra}[1]{{\langle#1|}}

\newcommand{\mod}[1]{{\rm mod}\left(#1\right)}
\begin{document}
\baselineskip 3.4ex
\begin{center}
{\huge \bf  The Los Alamos Trapped Ion Quantum Computer Experiment}\\
\bigskip\bigskip
{\small by }\\
\bigskip
{R. J. Hughes [1]\footnote{Address correspondence to
Dr. Richard J. Hughes, Physics Division P-23,  Mail Stop H-803, Los Alamos 
National Laboratory, Los Alamos NM 87545, USA.  Email 
hughes@lanl.gov.  More information
can be obtained at our website:
http://p23.lanl.gov/Quantum/quantum.html}, 
D. F. V. James [4], J. J. Gomez [3], M. S. Gulley [3],}\\ 
{ M. H. Holzscheiter [1], P. G. Kwiat [1], S. K. Lamoreaux [1], C. G. Peterson [1],}\\
{V. D. Sandberg [1], M. M. Schauer [2], C. M. Simmons [1],
C. E. Thorburn [1]\footnote{Current address: Engineering Sciences and 
Applications Division ESA-MT, MS C-914, Los Alamos National Laboratory},}\\
{D. Tupa [1], P. Z. Wang [1]\footnote{Permanent address: 
Physics Dept., Cornell University, Ithaca, N.Y.}, A. G. White [1].}\\
\bigskip
{\small [1] Physics Division P-23, Mail Stop H-803, } \\
{\small [2] Physics Division P-24, Mail Stop E-526, } \\
{\small [3] Physics Division P-25, Mail Stop H-846, } \\
{\small [4] Theoretical Division T-4, Mail Stop B-268, } \\
{\small University of California}\\
{\small Los Alamos National Laboratory}\\
{\small Los Alamos, NM87545, USA}\\
\bigskip
{submitted to \em {Fortschritte der Physik}}\\ 26 August 1997\\
\bigskip
\bigskip
{\large \bf Abstract}
\end{center}
The development and theory of an experiment to investigate quantum computation with
trapped calcium ions is described. The ion trap, laser and ion requirements
are determined, and the parameters required for quantum logic operations as
well as simple quantum factoring are described. 
\bigskip
\begin{center}
LA-UR-97-3301\\
\end{center}

\newpage
\section{Introduction}\setcounter{equation}{0}
In the early 1980's, various authors started to investigate the generalization
of information theory concepts to allow the representation of information by
quantum states. The introduction into computation of {\em quantum} physical
concepts, in particular the superposition principle, 
opened up the possibility of new capabilities, such as quantum
cryptography \cite{HughesCrypto}, that have no classical counterparts. One of the most
intriguing of these new ideas is quantum computation, first proposed by
Benioff \cite{Benioff}. Subsequently Feynman  \cite{Feynman} suggested
that quantum computation might be
more powerful than classical computation. This notion gained further
credence through the work of Deutsch who introduced the idea of quantum
parallelism to describe the ability of a quantum computer to compute with
quantum superpositions of numbers \cite{Deutsch}. 
Deutsch and Josza illustrated this
power of quantum computation with an Oracle problem \cite{DeutschJoza}.
However, until 1994
quantum computation was an essentially academic endeavor because there
were no quantum algorithms that exploited this power to 
solve interesting computational problems, and because no realistic
hardware capable of performing quantum computations had been envisioned. 
But then, building on earlier work of Simon \cite{Simon},
Shor discovered quantum
algorithms for efficient solution of two problems that are at the heart of
the security of much of modern public key cryptography: integer
factorization and the discrete logarithm problem 
\cite{Shor, EkertJozsa}. Later in 1994 
Cirac and Zoller proposed that quantum computational hardware could be
realized using known techniques in the laser manipulation of trapped 
ions \cite{CZ}.
Since then interest in quantum computation has grown dramatically, with
several groups, including our own, now pursuing trapped ion and other
quantum computation experiments.

Since 1994 remarkable progress has been made: a single quantum logic gate
has been demonstrated with trapped ions \cite{NISTgate};
quantum error correction schemes
have been invented \cite{MannyRay, Preskill};
and quantum algorithms for solving new problems have
been discovered \cite{Grover, Terhal, Boneh, Kitaev}. 
In this paper we will review the subject of quantum
computation and our development of an experiment using trapped calcium ions.
Also, we will explore the potential of this technology and determine whether
the future obstacles to progress will be fundamental or technical.

The remainder of this paper is organized as follows.  In section 2, 
the essential principles of quantum computation are introduced.  
Section 3 briefly describes the potential for efficiently factoring large
numbers with a quantum computer.  In section 4, we discuss various 
quantum computational hardware ideas; the trapped ion proposal is then
described in detail in section 5.  The experiment we are undertaking to
realize such a trapped ion quantum computer is described in
section 6, and the last section summarizes our conclusions.

\section{The Principles of Quantum Computation}\setcounter{equation}{0}

The principles of quantum computation have been discussed in detail 
elsewhere (see for example \cite{EkertJozsa, Lloyd})
so we shall only give a very brief
introduction here.  
The essential idea of quantum computation is to represent binary numbers 
using a collection of
two-level quantum systems. We will use the notation $\left\{ \ket{0}
,\ket{1} \right\} $ to denote the
states of a single two-level quantum system,
known as a quantum bit or {\em qubit}. 

With multiple qubits the number of degrees of freedom in
the system rapidly increases: the Hilbert space describing the
state of a system containing $N$ qubits has $2^{N}$ dimensions.
It is possible to use such a system to
represent a number, $x$, between 0 and $\left( 2^{N}-1\right) $, as the state
\be
\ket{x} \equiv \prod\limits_{i=0}^{N-1}\ket{x_i}_{i},
\ee
where $x=\sum\limits_{i=0}^{N-1}x_{i}2^{i}$, and $x_{i}$ is the $i$th 
binary digit of 
$x$.  Thus, for example, the decimal number 11, which is written as 1011 in 
binary, would be represented by four qubits in the state 
$\ket{1}_{3}\ket{0}_{2}\ket{1}_{1}\ket{1}_{0}$, which is more
conveniently written in the short-hand notation $\ket{1011}$.

To perform computations, we need to be able to perform
certain unitary operations that act on sets of qubits,
known as {\em quantum logic gates}. Because quantum interactions are
reversible, the underlying logic of a quantum computer must itself be
reversible. Fortunately, it is already known that arbitrary Boolean
operations can be constructed reversibly \cite{Bennet}, 
\cite{FredkinToffoli}.

 An example of a reversible logic
operation is the $NOT$ operation on a single qubit:
\be
NOT:\ket{b} \rightarrow \ket{\bar{b}} .
\ee
Arithmetic operations require logic operations between two or more qubits.
For example, in the controlled-NOT operation on two qubits
\be
CNOT:\ket{ c} \ket{ t} \rightarrow \ket{
c} \ket{ t\oplus c} ,
\ee
a target qubit, t, is flipped in value when a control qubit, 
c, has the value 1, but is unchanged when the control has the value 0. The symbol
$\oplus$ denotes addition modulo 2, defined by the following truth 
table:
\begin{center}
\begin{tabular}{|cc|c|}\hline
a&b&a $\oplus$ b\\ \hline
0&0&0\\
0&1&1\\
1&0&1\\
1&1&0\\ \hline
\end{tabular}.
\end{center}
A second application of
the CNOT gate returns the state to its starting value.

Another reversible logic gate used in
quantum computing is the  three-bit controlled controlled-NOT
(or Toffoli gate):
\be
CCNOT:\ket{c_1}\ket{c_2}\ket{t}\rightarrow 
\ket{c_1}\ket{c_2}\ket{\left( c_1 \wedge c_2 \right) \oplus t} ,
\ee
where $\wedge$ denotes the logical AND defined by
the following truth table:
\begin{center}
\begin{tabular}{|cc|c|}\hline
a&b&a $\wedge$ b\\ \hline
0&0&0\\
0&1&0\\
1&0&0\\
1&1&1\\ \hline
\end{tabular}.
\end{center}
Using the Toffoli and CNOT gates it is possible to construct 
a simple binary adder:
\be
ADD\left(\ket{a},\ket{b},\ket{0} \right) =
CNOT_{1,2}\left( CCNOT\left(\ket{a},\ket{b},\ket{0} \right) \right)
=\ket{a}\ket{a\oplus b} \ket{a\wedge b}
\ee
which places the sum modulo 2 of the first and second inputs (reading left to
right) onto the second output, and the carry bit (originally reading 0) into
the third output qubit. (Here, the CNOT-operation on three qubits is defined
as: $CNOT_{1,2}:\ket{a}\ket{b}\ket{c} \rightarrow 
\ket{a}\ket{a\oplus b}\ket{c} $ .)
With these {\em universal} quantum logic gates,
arbitrary arithmetic functions can be formed.

So far our discussion has not revealed any particular power associated
with quantum computation. Indeed, it is clear that the physical requirement
of logical reversiblity leads to extra information (relative to
conventional computation) being carried forward. This extra information,
which allows the input of a logical operation to be determined from the
output, imposes additional memory requirements on quantum computation.
Nevertheless, it is known that the peculiarly quantum concepts of
superposition and interference can be utilized to provide a much more
efficient solution of certain problems than is possible on any conventional
computer. We consider computations with superpositions of numbers first. 
A unitary operation on a single qubit can be written as
\be
\hV\left( k,\phi \right) :\left\{ 
\begin{array}{l}
\ket{0} \\ 
\ket{1}
\end{array}
\right\} \rightarrow \left\{ 
\begin{array}{l}
\cos \left( \frac{k\pi }{2}\right) \ket{0} -i\exp \left(i\phi
\right) \sin \left( \frac{k\pi }{2}\right)\ket{1} \\ 
\cos \left( \frac{k\pi }{2}\right) \ket{1} -i\exp \left(
-i\phi \right) \sin \left( \frac{k\pi }{2}\right) \ket{0}
\end{array}
\right\} .
\ee
For example, $\hV\left( 1/2,\pi /2\right) :\ket{ 0} \rightarrow
2^{-1/2}\left( \ket{ 0} +\ket{ 1} \right) $ , is
an equally-weighted superposition of both qubit states. Now, starting with
an $L$-qubit register in the state $\ket{ 0} $ $\equiv
\prod\limits_{i=0}^{L-1}\ket{ 0} _{i}$ , and acting on each
qubit with $\hV\left( 1/2,\pi /2\right) $ produces a coherent superposition of
all $2^{L}$ possible numbers:
\be
\ket{ 0} \rightarrow 2^{-L/2}\prod\limits_{i=0}^{L-1}\left(
\ket{ 0} _{i}+\ket{ 1} _{i}\right)
=2^{-L/2}\sum\limits_{a=0}^{2^{L}-1}\ket{ a}.
\ee
Thus, in a sense the memory of a quantum computer is exponentially large. 
Now the quantum
register state above is a product state, whereas a more typical
superposition state of the register will be an entangled state
(i.e. a state which cannot be written as a simple tensor product of 
basis states). Such states
can be produced with the aid of quantum logic operations. For example, an
entangled state of two qubits can be produced from an initial product state:
\be
CNOT:2^{-1/2}\left( \ket{ 0} +\ket{ 1} \right)
\ket{ 0} \rightarrow 2^{-1/2}\left( \ket{ 0}
\ket{ 0} +\ket{ 1} \ket{ 1} \right) .
\ee

Entangled states are required during typical quantum computations (and it is 
the existence of such entanglements that distinguish a quantum
computer from a classical computer), but because of
their non-classical properties the CNOT operation is extremely difficult to
construct in a physical system. 

Thus far we have seen the power of quantum
memory. Now we explore the power of actual quantum computations. Let us
suppose that we have determined the sequence of quantum gates required to
evaluate (reversibly) the value of some function, F, for an arbitrary input,
a:
\be
\hat{F}:\ket{ a} \ket{ 0} \rightarrow \ket{
a} \ket{ F\left( a\right) } ,
\ee
where the argument is held in the left register, and the right register is
to hold the function value. We might as easily have started with the left
register in an equally-weighted superposition of all values, and applied the
same sequence of logic gates:
\be
\hat{F}:\sum\limits_{a=0}^{2^{L}-1}\ket{ a} \ket{ 0}
\rightarrow \sum\limits_{a=0}^{2^{L}-1}\ket{ a} \ket{ F\left(
a\right) } ,
\ee
evaluating all $2^{L}$ function values in one step. 
(Recall that the initial state of
the left register can be created from the $\ket{ 0} $ state
with only $L$-single qubit unitary operations.) If we were only interested
in the function values we would now have to repeat the creation of this
state $O(2^{L})$ times, and measure the right register each time in order to
determine the values. Obviously, for this type of problem quantum
computation offers no advantages, but if we were instead interested in some
joint property shared by all the function values, such as the function's
period, we could now perform a quantum Fourier transform (QFT) operation on the
left register to determine the period efficiently. The point is that
particular function values in the right register are associated with
sequences of values in the left register that reflect the period. The
QFT, which is given by the following formula
\be
\ket{ a} \rightarrow 2^{-\frac{L}{2}}\sum
\limits_{c=0}^{2^{L}-1}\exp \left( i\frac{a.c}{2^{L}}\right) \ket{
c} ,
\ee
regroups these sequences to produce constructive interference at values
corresponding to periods of the sequence. Furthermore, the QFT itself can be
constructed using only $O(L^{2})$ quantum gates, whereas a conventional
discrete Fourier transform requires $O(L2^{L})$ operations for a 
$L$-bit input register. For problems
that can be reduced to determining the period of a function, quantum
computation may offer a more efficient solution than conventional
computation. All known quantum algorithms to solve interesting problems use
either the QFT or one of its variants, such as the quantum Hadamard
transformation.

\section{Quantum Factoring}\setcounter{equation}{0}

The power of quantum computing can be illustrated with the example of Shor's
algorithm for integer factorization.
According to the fundamental theorem of arithmetic, every integer has a
unique expression as a product of primes. Most modern factoring algorithms
use as their starting point the Legendre congruence: given a composite
integer, N, which we want to factor, the congruence
\be 
y^{2}=1\,\mod{N}
\ee
has non-trivial solutions, $y=\pm a\,\mod{N}$, in addition to the trivial
solutions, $y=\pm 1\,\mod{N}$. If a non-trivial solution, $a$, can be found,
it can be used to find a factor of N, because the congruence
\be
\left( a+1\right) \left( a-1\right) =0\,\mod{N}
\ee
implies that the factors of $N$ are distributed between the two parentheses.
Therefore, the greatest common divisor ($gcd$)
\be
gcd (a\pm 1,N) = \mbox{factor of N},
\ee
which can be found using Euclid's algorithm, gives a factor of $N$. The
problem of integer factorization can therefore be reduced to finding a
suitable a.

Shor's algorithm for finding factors using a quantum computer
starts by using a classical computer to find an 
integer x which shares no common factors other than 1 with N:
\be
\left( x,N\right) =1
\ee
Once a value of $x$ is known the function given by the following 
expression can be defined:
\be
f_{x}\left( z\right) =x^{z}\,\mod{N} , \: z=0,1,2,\cdots, N^2-1 .
\ee
This function will be periodic, with some period r, (known as the order of $x
\,\mod{N}$).  Mathematically, for some r,
\be
f_{x}\left( z+r\right) =f_{x}\left( z\right),
\ee
and hence
\be
x^{r}=1\,\mod{N}.
\ee

Thus, if r is even $a=x^{r/2}\,\mod{N}$ is a candidate solution of
Legendre's congruence and hence allows factorization of N. (There are some
technicalities about the choice of the upper limit to the domain of 
$f_{x}$ and the
classical post-processing required to deduce r from the quantum part of the
algorithm.)

Shor's algorithm requires a $2\log _{2}N$ bit left quantum register to hold
the argument z, and an $l=$ $\log _{2}N$ bit right register to hold the
value of the function f:
\be
\ket{ \psi } =\prod\limits_{i=0}^{2l-1}\ket{ 0}
_{iL}\prod\limits_{j=0}^{l-1}\ket{ 0} _{jR}\equiv \ket{
0} _{L}\ket{ 0} _{R}
\ee
where the suffix i (j) denotes the i-th (j-th) qubit of the left (right) register
respectively.

Next, the left register is prepared in a superposition of all possible 
values,  $z$,
using 2l operations, $\hV \left( 1/2,\pi /2\right) $, one for each qubit:
\be
\ket{ \psi } \rightarrow \prod\limits_{i=0}^{2^{2l}-1}\hV_{i}\left(
1/2,\pi /2\right) \ket{ \psi } =\frac{1}{2^{l}}
\sum\limits_{i=0}^{2l-1}\ket{ z} _{L}\ket{ 0}
_{R}\equiv \ket{ \psi _{1}}
\ee

In the next step the function $f_{x}\left( z\right) $ is written into the
right register:
\be
\ket{ \psi _{1}} \rightarrow \frac{1}{2^{l}}
\sum\limits_{i=0}^{2^{2l}-1}\ket{ z} _{L}\ket{ f_{x}\left(
z\right) } _{R}\equiv \ket{ \psi _{2}} 
\ee

Because the function $f_{x}$ is an exponential, the computation to build the
function can be broken down into repeated multiplications, which can be
reduced to repeated addition, and ultimately represented in terms of the
simple adder and similar circuits. The total number of quantum gates
required is therefore $O\left( l^{3}\right)$ and some additional register
space (roughly $2l$ qubits) is required for this computation to proceed
reversibly.  The resulting state has the form
\be
\ket{ \psi _{2}} \sim \frac{1}{2^{l}}\sum\limits_{z_{0}}
\sum_{j}\ket{ z_{0}+jr} _{L}\ket{ f_{x}\left( z_{0}\right)
} _{R}
\ee
in which sequences of left register states, separated by the period r,
are associated with common right register states. At this stage, it
is easier to think about making a measurement of the right register,
resulting in a particular value $f\left( z_{0}\right) $ :
\be
\ket{ \psi _{2}} \rightarrow \ket{ \phi }
\sim \sum\limits_{j}\ket{ z_{0}+jr} _{L}.
\ee

Finally, a quantum Fourier transform can be performed on this state after
which a measurement is made, resulting in some value $2^{2l}/jr$ for some j.
From this result, the order r can be determined, and hence N can be
factored. Note that there are some subtleties involved when r does not
divide $2^{2l}$ , and also the algorithm must be repeated until an
$x$ with an
even r is found.  

To put Shor's factoring algorithm in perspective, the total number of logic 
gates, $N_{g}$ and the total number of
qubits $L$ required to factor an $\ell$-bit integer using
a version of the algorithm that we have recently developed are
\cite{HughesNeergard}:
\bea
N_{g}&=&24\ell^{3}+ O(\ell^{2}) \nn \\
L&=&5\ell + 4.
\label{quantfac}
\eea
In contrast, the best algorithm for factoring a large
integers $N$ using classical computers is the (general) Number
Field Sieve (NFS) \cite{Lenstra}, which runs in
an asymptotic heuristic time $T_{R}$ given by:
\be
T_{R} \sim \exp\left[1.923\ell^{1/3} (\log \ell)^{2/3}\right].
\label{classfac}
\ee
The NFS has much faster growth with the number of bits in N than the 
polynomial growth of the quantum algorithm. The NFS was recently used to 
factor the 130-digit (430-bit) integer known as RSA130 
\cite{Cowie} using 
approximately 500 MIPS-years of computer time distributed over the 
Internet. (1 MIPS-year is approximately $3.0\times10^{13}$ instructions.) 
With one 
hundred machines each rated at 100 MIPS dedicated to this problem we would 
estimate a factoring time of approximately 18 days. In contrast the quantum 
factoring algorithm applied to the same integer would require only  
$\sim 2.0\times 10^{9}$ quantum logic gates. 
If a quantum computer had a clock speed of 100 
MHz it would factor this number in only $\sim$20 seconds. Furthermore, the 
tremendous efficiency of the quantum computer would become even more 
pronounced with larger numbers.


\section {Quantum Computer Technologies}\setcounter{equation}{0}
The three essential requirements for quantum computational hardware are: 
(1) the ability to isolate a set of two-level quantum systems from 
the environment for long enough to maintain 
coherence throughout the computation, while at the same time
being able to interact with the systems strongly enough to
manipulate them into an arbitrary quantum state; 
(2) a mechanism for performing quantum logic operations: in
other words a ``quantum bus channel'' connecting the various
two-level systems in a quantum mechanical manner; and
(3) a method for reading out the quantum state of the system
at the end of the calculation.

So far the serious proposals for quantum
computational hardware fall into five basic categories:

\noindent 
1. $\underline{{\rm Ion\,\,traps}}$. In this scheme the quantum data registers 
consist of internal
levels of ions trapped in a linear configuration; the 
quantum bus channel is realized using the phonon modes of the ions'
collective oscillations; and readout is performed by
using quantum jumps.  This scheme was originally proposed 
by Cirac and Zoller
in 1994 \cite{CZ}, and was used to demonstrate a 
CNOT gate shortly afterwards \cite{NISTgate}.
The number of qubits that can be realized using
an ion trap is limited by the various decoherence
mechanisms, which have been discussed in detail
elsewhere \cite{HJKLP, PK, HR, HJKLPtwo},
and by the onset of instabilities in the linear
configuration of the ions \cite {Schiffer}.
One estimate \cite{HJKLP} suggests
that computations involving 47 qubits and $4.0\times 10^{5}$
operations may be possible with trapped Calcium ions before 
decoherence becomes a serious problem, although this
estimate ignores experimental decoherence effects such as
ion heating and fluctuations of laser phases, as well as recent 
advances in the field of quantum error correction.  
(A variant of this idea using a photon mode of
a cavity as the quantum bus bit has also been proposed \cite{PGCZ}.)  
Our progress towards realizing a trapped ion quantum computer is
the subject of the remainder of this article.

\noindent 
2. $\underline{{\rm Cavity\,\,Quantum\,\,Electrodynamics}}$ (Cavity QED). 
Here the quantum data 
register consists of the photon modes of an optical cavity, which
are linked to other modes via excitation of, and emission by, an atom
passing through the cavity.  This scheme is a development
of experiments in cavity quantum electrodynamics, and it has been
used successfully to realize quantum gates, although scaling up to 
more than 2 or 3 qubits will probably be very difficult 
\cite{Kimble, Haroche}. A related {\em all photon} cavity QED realization has
been proposed \cite{Sten}.

\noindent
3. $\underline{{\rm Nuclear\,Magnetic\,Resonance}}$ (NMR).
The orientation of spin-1/2 nuclei
in a molecule form the data register, and the spin-spin interactions
provide the quantum bus channel.  The states of the spins can be altered 
by applying a radio-frequency magnetic field and 
readout is performed by measuring
the magnetization of bulk samples, a procedure which
can only measure an {\em ensemble average} of the quantum state
populations.  There are considerable subtleties associated with
preparation of the initial states as well as with the readout;
for this reason, the NMR scheme has been called ``{\em ensemble
quantum computing}'' \cite{Havel2} (see also \cite{Gershenfeldt}).
This method has been used to realize experimental quantum gates
involving two and three qubits
\cite{Havel1}, however because the signal falls exponentially
with the number of qubits, the ability to scale up to large
numbers of qubits seems to be problematic \cite{Warren}: 
the sources cited above give estimates
of the largest number of qubits that can be realized with NMR
computers that vary from 6 to 20. 

Similar concepts involving
electron spins \cite{Ziolo}, the use of atomic force microscopy
to manipulate nuclear spins
\cite {DiVincenzo} and the interactions
between electron and nuclear spins in the quantum Hall regime
\cite {Privman} have been proposed.

\noindent
4. $\underline{{\rm Superconducting\,\,Quantum\,\,
Interference\,\,Devices}}$ (SQUIDS).  In
this proposed realization, which is still being
developed and has not yet been used to realize a quantum logic
gate, the quantization of flux in a superconducting circuit
would be exploited to give two level systems \cite{Lukens, Bocko, Rosen}.
The principal attraction of this proposal
is that it is a solid state device, and so if it can be made
to work with a small number of qubits, scaling up to large numbers
of qubits should be relatively straightforward because the
technology for making miniaturized solid state devices containing
a very large number of computing elements already exists.  
However, solid state devices will also have their drawbacks, in that
they will be strongly coupled to a complex environment, so 
decoherence may become an insurmountable problem.

\noindent 
5. $\underline{{\rm Quantum\,\,Dots}}$.  Quantum dots are regions of 
artificial inhomogeneity
in a crystal, which can be placed in a controlled manner.  They can
be used to trap single electrons, the orientation of whose spin
can then be used as to store information \cite{Ekert, DiVincenzo2}.
Like the SQUID proposal, this idea has not yet been used to
demonstrate a logic gate, but offers the same advantage of
being a solid state device.


\section{Theory of Quantum Computation with 
Ions in a Linear Trap}\setcounter{equation}{0}

We shall now describe in detail the Cirac-Zoller trapped-ion 
quantum computer concept \cite{CZ} (see also refs. \cite{Steane, NISTrev}). 
In this scheme, illustrated in fig.1,  each qubit is the
electronic ground state ($\ket{ 0} $) and a metastable excited
state ($\ket{ 1} )$ of an ion, which has been laser cooled to
rest in a linear radio-frequency quadrupole ion trap. Computational
operations are performed with coherent laser-ion interactions driving Rabi
oscillations between the relevant quantum states of the register. At the end
of the computation, the results are read out using the quantum jump 
technique. The principal
advantages of the trapped ion scheme are that
many of the techniques required
to prepare and manipulate quantum states have already been
developed for precision spectroscopy work; 
secondly that the decoherence rates owing to
decay of the excited ionic state and heating of the ionic motion can be
made small in the virtually perturbation-free (ultra-high vacuum, low
electromagnetic noise level) environment of an ion trap; and finally there is an
experimentally demonstrated technique, quantum jumps, for reading out
the result of a computation with high probability.

The first step that must be accomplished for quantum computation is to
prepare an isolated quantum state of several ions in which each ion is
spatially localized and cooled to the quantum ground state of its 
vibrational motion.  The techniques employed for trapping and cooling 
the ions are described in
detail below.  
Operating the trap at low ion densities, Doppler cooling produces a so-called
``string-of-pearls'' configuration: the ions become arranged in a line 
and each is localized in a spatial region whose dimensions 
are very much less than the
wavelength of the optical transition. The spacing between 
adjacent ions can be made
conveniently large (30 $\mu$m, say) by striking a balance between their
Coulomb repulsion and the axial confining potential. This crystallized
structure permits the ions to be addressed by lasers for quantum computation.
It has been demonstrated that
a single ion can be cooled to its vibrational ground state in an ion trap
using resolved sideband cooling in the optical regime \cite{NISTcool}. 
Because we shall employ the quanta of the ions' collective vibrations
(i.e. phonons) as a 
``quantum bus channel'' connecting different qubits, these vibrations 
must initially be in their ground states.
However, cooling to
the vibrational ground state has yet to be demonstrated for two or more
ions. With the ions in their internal and phonon ground states and with
the cooling lasers turned off the system is prepared for the actual quantum
computation to begin.

\subsection{Phonon Modes}
We shall assume that ions are sufficiently tightly bound
in the directions transverse to the trap axis that we need only
consider their motion along the axial direction (i.e. the x-axis in fig.1).
Suppose that the m-th ion is displaced a small distance
$q_{m}$ from its equilibrium position $x^{0}_{m}$.  The Lagrangian
for the oscillations of the ions about their equilibrium
positions is
\be
L\left(\dot{q}_m,q_m\right) =\frac{M}{2}\sum^N_{m=1}(\dot{q}_m)^2
-\frac{1}{2}\sum^N_{n,m=1}C_{nm}q_{n}q_m,
\label{lag1}
\ee
where $N$ is the number of ions, $M$ is the mass of each ion
and the coupling matrix $C_{nm}$ is defined by
\be 
C_{nm}=\left[\frac{\partial^2V}{\partial x_n \partial x_m}\right]_0,
\ee
where $V$ here stands for the ions' potential energy, and
the subscript $0$ denotes that the double partial derivative
is evaluated at $q_n=q_m=0$. The potential energy, which consists
of two parts, the binding potential due to the trap
electrodes and the Coulomb interaction energy between the
ions themselves, is given by:
\be
V\left(x_{m}\right)=\sum^N_{m=1}\frac{1}{2}M\omega^2_{x}x_m(t)^2+
\sum_{\stackrel{\scriptstyle n,m=1}{m \neq n}}^N
\frac{e^2}{8\pi\epsilon_0}\frac{1}{|x_n(t)-x_m(t)|} \ ,
\label{potty}
\ee
where $ x_m(t)=x^{0}_{m}+q_{n}(t)$ is the position of the
m-th ion, $e$ is the charge of the electron (we have
assumed that the ions are singly ionized), $\epsilon_0$
is the permittivity of free space and 
$\omega_{x}$ is the angular frequency characterizing the
the harmonic binding potential.  If there were only one ion
confined in the trapping potential, then its oscillations would
be harmonic with this angular frequency. 
The elements of the matrix $C_{nm}$ may be calculated by 
differentiating eq.(\ref{potty}) and evaluated using the 
values for the equilibrium positions of the ions,
obtained numerically (see ref. \cite{dfvj}).  

The eigenvectors $b^{(p)}_m$ $(p=1,2,\ldots N)$ of the matrix $C_{nm}$
are defined by the following formulas:
\be
\sum^N_{n=1}C_{nm}b^{(p)}_n=\mu^{2}_{p}b^{(p)}_m 
\mbox{  } (p=1,\ldots, N) \ ,
\ee
where $\mu^{2}_p\;(p=1,2,\ldots N)$ are the eigenvalues (which are 
always positive).
The eigenvectors  are assumed to
be numbered in order of increasing eigenvalue and to be 
properly normalized. The first eigenvector 
(i.e. the eigenvector with the smallest eigenvalue)
can be shown to be
\be
\bb^{(1)}=\frac{1}{\sqrt{N}}\left\{1,1,\ldots,1\right\}{\rm ,}\qquad \mu_1=1 \ 
\label{beeone}. 
\ee

The {\em normal modes} of the ion motion are defined 
by the formula
\be
Q_p(t)=\sum^N_{m=1}b^{(p)}_{m}q_m(t)\;(p=1,2,\ldots N) \ .
\ee
The first mode $Q_1(t)$ corresponds to all of the
ions oscillating back and forth as if they were rigidly
clamped together;  this is referred to as the {\em center of mass}
(CM) mode.  The second mode $Q_2(t)$ corresponds to each
ion oscillating with an amplitude proportional to its
equilibrium distance from the trap center.  This is called
the {\em breathing mode}.  There are a total of $N$ modes
altogether.

The Lagrangian for the ion oscillations, eq. (\ref{lag1}),
may be rewritten in terms of these normal modes as follows:
\be
L\left(\dot{q}_m,q_m\right)=\frac{M}{2}\sum^N_{p=1}\left[\dot{Q}^2_p
-\mu^{2}_p\omega_{x}^2Q^2_p\right] \ .
\ee
Thus the canonical momentum conjugate to
$Q_p$ is $P_p\equiv\partial L/\partial \dot{Q}_p = M\dot{Q}_p$, 
and we can 
write the Hamiltonian as
\be
H\left(p_m,q_m\right)=\frac{1}{2M}\sum^N_{p=1}P^2_p+
\frac{M}{2}\sum^N_{p=1}(\omega_{x}\mu_p)^2 Q^2_p \ .
\ee
The quantum motion of the ions can now be considered
by introducing the operators:
\bea
Q_p\rightarrow\hat{Q}_p&=&i\sqrt{\frac{\quad\hbar\quad}
{2M\omega_{x}\mu_p}}
\left(\ha_p-\had_p\right) \ , \\
P_p\rightarrow\hat{P}_p&=&\sqrt{\frac{\hbar M\omega_{x}\mu_p}{2}}
\left(\ha_p+\had_p\right) \ .
\eea
where $\hat{Q}_p$ and $\hat{P}_p$ obey the canonical commutation
relation $\left[\hat{Q}_p,\hat{P}_q\right]=i\hbar\delta_{pq} $ and 
the creation and 
annihilation operators $\had_p$ and $\ha_p$ obey the
usual commutation relation 
$\left[\ha_p,\had_q\right]=\delta_{pq} $. 
  
We shall use the interaction picture to perform
calculations of the effect of laser fields on the trapped ions.
The unperturbed Hamiltonian describes the free evolution of
the internal states of the ions and the oscillations of the ions'
normal modes.  The effect of the laser field will be described by
an interaction Hamiltonian introduced below.  The interaction picture
operator for the displacement of the 
m-th ion from its equilibrium position is:
\bea
{\hat q}_m(t)&=&\sum_{p=1}^N b^{(p)}_m \hat{Q}_p(t)\\
&=& i\sqrt{\frac{\hbar}{2M\omega_{x}N}}
\sum^N_{p=1}s^{(p)}_m
\left(\ha_{p}e^{-i\omega_{x}\mu_{p}t}-\had_{p}e^{i\omega_{x}\mu_{p}t}\right) \ , 
\label{deltaxhat}
\eea
where the coupling constant is defined by
\be
s^{(p)}_m=b^{(p)}_m \sqrt{N/\mu_p}\ .
\label{coupconsts}
\ee
For the CM mode (p=1), which is the most important mode for
quantum computation, $s^{(1)}_m=1$ for all ions $m=1,2,\ldots N$.

\subsection{Laser-ion interactions}
\subsubsection{``V'' Type Operations: Single Qubit Interactions}
There are two different ways of performing quantum 
computational laser operations on
the qubits formed from two-level ions.  The most
simple is the single frequency Rabi oscillations between two states.
If the two level system interacts
with a monochromatic field precisely tuned to the transition
frequency for the two levels, the population will oscillate back
and forth between the two levels (fig. 3.a)
\cite {FeynmanVernonHellwarth, AllenEberly}.  A long-lived
two level system is required to make a qubit suitable for 
quantum computation, and so atomic levels with
dipole-forbidden transitions are the most suitable.  
An example of such a dipole forbidden transition is the
729 nm $4\,^{2}S_{1/2}$ to $3\,^{2}D_{5/2}$ transition in $Ca^{+}$ 
(fig.2), 
which has a natural lifetime of about 1.06 seconds
(see ref.\cite{dfvj} for more details and references to the
sources of atomic data).

The other method of performing manipulations on the qubits
of an ion trap quantum computer is to use Raman transitions (fig.3.b).
This technique was used by the Boulder group
in their experimental realization of a quantum logic gate
using a cold trapped ion \cite{NISTgate}.
Two lasers, traditionally named the ``pump'' and the ``Stokes''
beams, are tuned so that population from one level is
pumped to some intermediate virtual level by one laser, 
and then immediately
brought back down from the virtual level to a different
atomic state by the second laser \cite{SchenzleBrewer, dfvj2}.
This method has some advantages over the single
laser technique, because it is resilient against phase fluctuations
of the addressing laser and because one can use two states with
{\em very} long decay times, for example the two magnetic sub-levels of
the $4\,^{2}S_{1/2}$ level of $Ca^{+}$.  However it also has the 
complication that one needs to precisely control the position and
direction of two
laser beams rather than the one required for the first method.  
It may be that for the experimental devices now under construction,
which will consist of only a few qubits, the first technique will
prove superior.

The interaction picture Hamiltonian describing the resonant interaction
of a two level system with a laser can be written as follows:
\be
\hat{H}_{I}=\frac{i\hbar\Omega_{0}}{2}\exp{(i\Delta 
t)}\ket{0}\bra{1}+h.a.\label{Hint}
\ee
where $h.a.$ stands for the Hermitian adjoint of
the immediately preceding term.
The Rabi frequency $\Omega_{0}$ is defined by
\be
\Omega_{0}=\left\{ 
\begin{array}{ll}
\frac{\D e E}{\D \hbar} 
\sqrt{\frac{\D A}{\D c\alpha k^{3}}}
\beta& \mbox{(single laser)}\\
&\\
\frac{\D e^{2} A }{\D \hbar^{2}c\alpha k^{3}}
\frac{\D E_{p}E_{s}^{\ast}}{\D 4\delta}\beta& \mbox{(Raman)}
\end{array}
\right.
\label{Omega0}
\ee
In these formulas, $\beta$ is
a coefficient of order 1 which depends on the laser polarizations
and the quantum numbers of the states involved (for details see
\cite{dfvj, dfvj2}); 
 $E$ is the complex electric field strength
for the single laser; $E_{p}$ and $E_{s}$ are the complex 
electric field strengths of the pump and Stokes lasers respectively;
$A$ and $k$ are, in the single laser case,
the Einstein A coefficient and the wavenumber for the qubit transition, and in
the Raman case, the A coefficient and the wavenumber for the transition
from the two qubit levels $\ket{0}$ and $\ket{1}$ to the third detuned 
level,
$\ket{2}$ (see fig.3) \footnote{If $\ket{0}$ and $\ket{1}$ are 
magnetic sublevels of the same level, then the value of $A$ will be
the same for them both.};
$\delta$ is the detuning of the virtual level from level $\ket{2}$.
The detuning, $\Delta$, which appears in eq.(\ref{Hint}) is given by the 
following formula:
\be
\Delta=\left\{
\begin{array}{ll}
\omega_{0}-\omega_{L}&\mbox{single laser}\\
&\\
\omega_{0}-(\omega_{p}-\omega_{s})&\mbox{Raman}
\end{array}
\right.
\ee
where $\omega_{0}$ is the angular transition frequency for the
$\ket{0}$ to $\ket{1}$ transition, $\omega_{L}$ is the angular
laser frequency (single laser case) and $\omega_{p}$ and $\omega_{s}$
are the pump and Stokes angular frequencies respectively (Raman case)
\footnote{We have ignored the A.C. Stark shifts of the two 
qubit levels caused by the two Raman lasers.  This effect
will mean that the on-resonance condition will not be exactly
$\Delta=0$, and there will be a small phase shift in the execution
of CNOT gates as a result.  This is discussed in detail in 
ref.\cite{dfvj2}}.

If we assume that the detuning in eq.(\ref{Hint}) is zero, and that
only the states $\ket{0}$ and $\ket{1}$ can become populated,
i.e. the wavefunction for the qubit is $\ket{\psi 
(t)}=a(t)\ket{0}+b(t)\ket{1}$, then the equation of motion,
\be
i\hbar\frac{\partial}{\partial t}\ket{\psi(t)}=
\hat{H}_{I}\ket{\psi(t)} ,
\label{SE}
\ee
has the following solution:
\be\left(\begin{array}{c}
a(t)\\
b(t)
\end{array}\right)=\left(\begin{array}{cc}
\cos(\theta/2)            &          i e^{i\phi}\sin(\theta/2)\\
i e^{-i\phi}\sin(\theta/2) &          \cos(\theta/2)            
\end{array} \right)\left(\begin{array}{c}
a(0)\\
b(0)
\end{array}\right)
\ee
where $\theta=|\Omega_{0}|t$ and $\phi=\arg\{\Omega_{0}\}$.

Thus, by directing an on-resonance laser pulse at a
 particular ion, an
arbitrary superposition of the $\ket{0} $ and $\ket{1} $
states can be created from the $\ket{ 0} $ state by carefully
controlling the duration and phase of the pulse. The effect of such a  pulse,
which we will refer to as a ``V'' pulse,
on the m-th ion's internal states is represented by the single-ion unitary
operation, $\hat{V}_{m}\left(\theta,\phi \right) $,
\be
\hat{V}_{m}\left(\theta,\phi \right)  : 
\begin{array}{lll}
\ket{0}_{m}&\rightarrow&
\cos \left(\theta/2\right)\ket{ 0}_{m}
-i e^{i\phi }\sin \left( \theta/2\right)\ket{1}_{m} \\ 
\ket{1}_{m}&\rightarrow&
\cos \left( \theta/2\right)\ket{1}_{m}
-i e^{-i\phi}\sin \left(\theta/2\right)\ket{0}_{m}
\end{array} .
\ee
For example, $\theta=\pi$ is a $\pi $-pulse, which changes $\ket{
0} $ into $\ket{ 1} $ and vice-versa (with a $\pi $
phase shift):
\be
\hat{V}_{m}\left( \pi,\pi/2\right) : 
\begin{array}{lll}
\ket{0}_{m}&\rightarrow& \ket{1}_{m} \\ 
\ket{1}_{m}&\rightarrow&-\ket{0}_{m}
\end{array}
\ee
Similarly, the operation $\hat{V}\left(\pi/2,\pi/2\right) $, a $\pi /2$ pulse,
creates equally-weighted superpositions of the basis states; A 2$\pi$ pulse 
produces a sign change of the state in the same way that a
rotation of 2$\pi $ produces a sign change of the wavefunction of a spin-1/2
particle; a $4\pi$ pulse returns the ion to its original state.

However, these operations all have the form of a rotation,
whereas quantum logical operations are required that have the form of a
reflection, such as the NOT operation:
\be
NOT:
\begin{array}{lll}
\ket{0}_{m}&\rightarrow&\ket{1}_{m} \\ 
\ket{1}_{m}&\rightarrow&\ket{0}_{m}
\end{array}
\ee
and the single-bit Hadamard operation:
\be
\hat{R}: 
\begin{array}{lll}
\ket{ 0} _{m}&\rightarrow&\left( \ket{0}
_{m}+\ket{1} _{m}\right)/\sqrt{2} \\ 
\ket{1}_{m}&\rightarrow&\left(\ket{0}
_{m}-\ket{ 1}_{m}\right)/\sqrt{2}.
\end{array}
\ee
We can perform these operations 
with the aid of an {\em auxiliary level}, 
$\ket{aux}$.  The auxiliary level, which is
a vital ingredient of Cirac and Zoller's scheme,
is some third state of the ion
which will interact with $\ket{0}$ in precisely the
same manner as $\ket{0}$ interacts with $\ket{1}$, but which does not
interact at all with $\ket{1}$.  For example, 
we can use the $4^{2}S_{1/2},\,M_{J}=1/2$ and the
$3^{2}D_{5/2},\,M_{J}=3/2$ states of $Ca^{+}$ as $\ket{0}$
and $\ket{1}$ for our qubits.  The auxiliary level can then be
the $3^{2}D_{5/2},\,M_{J}=-1/2$ state.  We can perform operations
between $\ket{0}$ and $\ket{1}$ using right-hand circularly
polarized light.  If we switch to left circularly polarized light
we can perform operations between $\ket{0}$ and $\ket{aux}$, 
without changing the state of $\ket{1}$.
The unitary 
operations involving $\ket{0}$ and $\ket{aux}$ will be 
denoted as follows:
\be
\hat{V}^{(aux)}_{m}\left(\theta,\phi\right) : 
\begin{array}{lll}
\ket{0}_{m}&\rightarrow& 
\cos \left( \theta/2 \right)\ket{0}_{m}
-i e^{i\phi}\sin \left(\theta/2 \right)\ket{aux}_{m} \\ 
\ket{aux} _{m}&\rightarrow& 
\cos \left( \theta/2\right)\ket{aux}_{m}
-i e^{-i\phi}\sin\left(\theta/2 \right) \ket{0}_{m}
\end{array}
\ee
Using this operation $\hat{V}^{(aux)}_{m}$ in conjunction with
$\hat{V}_{m}$ , we can perform the NOT and single bit Hadamard 
transforms on the m-th qubit as follows:
\bea
NOT_{m}&=&\hat{V}^{(aux)}_{m}\left(2\pi,\pi/2\right) \hat{V}_{m}\left(\pi/2,\pi 
/2\right)\\
\hat{R}_{m}&=&\hat{V}^{(aux)}_{m}\left(2\pi,\pi 
/2\right)\hat{V}_{m}\left(3\pi/2,\pi/2\right)  .
\eea

\subsubsection{``U'' Type Operations: Interactions with the Quantum 
Bus Channel}

To describe the ability of the laser beam to excite or annihilate
quanta of the ions' collective vibrations, it is necessary to
modify the interaction Hamiltonian (\ref{Hint}) by taking into
account the spatial variation of the the laser field as follows:
\bea
E\rightarrow 
E(\hat{\bf{r}}_{m})&=&E({\bf r}^{0}_{m}+{\bf e}_{x}\hat{q}_{m})\nn \\
&\approx& E_{eq}+\hat{q}_{m}\left(\frac{\partial E}{\partial 
x}\right)_{0}
\eea
where ${\bf r}^{0}_{m}$ is the equilibrium position of the m-th ion,
${\bf e}_{x}$ is the unit vector in the axial (x) direction, $\hat{q}_{m}$
is the quantum operator describing the vibration of the m-th ion  
(see \ref{deltaxhat}) and the subscript $0$ denotes a term which has been 
evaluated at the equilibrium position of the ion.  For simplicity, we
shall neglect all of the phonon modes except for the CM mode (p=1).
We therefore obtain the following modified form of the interaction
Hamiltonian:
\bea
\hat{H}_{I}&=&\frac{i\hbar\Omega_{0}}{2}\exp{(i\Delta t)}
\ket{0}\bra{1}\nn \\
&+&\frac{i\hbar\Omega_{1}}{2}\exp{(i\Delta t)}
\ket{0}\bra{1}\left[\hat{a}\exp(-i\omega_{x}t)-
\hat{a}^{\dagger}\exp(i\omega_{x}t)\right]
+h.a.
\label{Hint2}
\eea
Here $\Omega_{0}$ is given by eq.(\ref{Omega0}), and
\be
\Omega_{1}=\sqrt{\frac{\hbar}{2M\omega_{x}N}} 
\left(\frac{\partial \Omega_{0}}{\partial 
x}\right)_{eq} ,
\label{Omega1}
\ee
where the partial derivative acts only on the spatial variations of 
the electric fields.  For plane waves, which will be a good 
approximation
for the focal regions of Gaussian laser beams in the vicinity of
an ion, $\Omega_{1}$ is given by the formula
\be
\Omega_{1}=\frac{\eta}{\sqrt{N}}\Omega_{0} ,
\label{Oemga1}
\ee
where $\eta$ is the Lamb-Dicke parameter defined by:
\be
\eta=\left\{\begin{array}{ll}
\sqrt{\frac{\D \hbar}{\D 2M\omega_{x}}}{\bf k}_{L}\cdot{\bf e}_{x}&
\mbox{(single laser)}\\
&\\
\sqrt{\D \frac{\hbar}{\D 2M\omega_{x}}}({\bf k}_{p}-{\bf k}_{s})\cdot{\bf e}_{x}&
\mbox{(Raman)}\\
\end{array} 
\right.
\ee
${\bf k}_{L}$, ${\bf k}_{p}$ and ${\bf k}_{s}$ being, respectively the
wavevectors of the single laser, the pump laser and the Stokes laser.
\footnote{The expression (\ref{Omega0}) is an approximation which
is valid if $\eta/\sqrt{N}\ll 1$.  In general the expression for
$\Omega_{0}$ depends on the number of phonons in the system; Monroe 
{\it et al.} recently pointed out that this fact allows one, in principal,
to realize quantum gates more simply than by using the Cirac and
Zoller auxiliary level scheme \cite{NISTsimp}.}

In order to realize quantum logic gates of the type devised by
Cirac and Zoller, one sets the detuning, $\Delta=-\omega_{x}$.
In this case, ignoring the off-resonant terms, eq. (\ref{Hint2}) 
becomes
\be
\hat{H}_{I}=\frac{-i\hbar\Omega_{1}}{2}\hat{a}^{\dagger}\ket{0}\bra{1}
+h.a.
\label{Hint3} 
\ee
This Hamiltonian is equivalent to that introduced by Jaynes and
Cummings to describe the interaction of a two-level system with
a single quantized harmonic oscillator \cite{JaynesCummings}. 
If we assume that we are limited to two possible states: 
$\ket{1}\otimes\ket{\mbox{0 phonons}}$ and
$\ket{0}\otimes\ket{\mbox{1 phonon}}$, which we shall
denote as $\ket{1,0}$ and $\ket{0,1}$ respectively,
the wavefunction for the ion is 
$\ket{\psi'(t)}=c(t)\ket{0,1}+d(t)\ket{1,0}$.  The
equation of motion (\ref{SE}) can then be solved:
\be\left(\begin{array}{c}
c(t)\\
d(t)
\end{array}\right)=\left(\begin{array}{cc}
\cos(\theta'/2)            &          i e^{i\phi'}\sin(\theta'/2)\\
i e^{-i\phi'}\sin(\theta'/2) &          \cos(\theta'/2)            
\end{array} \right)\left(\begin{array}{c}
c(0)\\
d(0)
\end{array}\right)
\ee
where $\theta'=|\Omega_{1}|t$ and  $\phi'=\arg\{\Omega_{1}\}$.
Thus we have interactions which change the CM phonon 
vibrational mode as well as the ions'
internal levels in a controllable manner. 
We shall refer to such laser interactions as ``U'' type pulses.  
The following operation
can be realized by these interactions with
precise control of the pulse duration and laser phase:
\be
\hat{U}_{m}\left(\theta',\phi' \right) : 
\begin{array}{lll}
\ket{0,1}_{m}&\rightarrow&\cos\left( 
\theta'/2\right) \ket{0,1}_{m}
-i e^{i\phi'}\sin\left(\theta'/2\right) \ket{1,0}_{m}\\ 
\ket{1,0}_{m}&\rightarrow&\cos \left( 
\theta'/2 \right) \ket{1,0}_{m}
-i e^{-i\phi'}\sin \left( \theta'/2 \right) \ket{0,1}_{m}
\end{array}
\ee
A similar transform $\hat{U}^{aux}_{m}$ can be defined between
the $\ket{0,1}$ state and the $\ket{aux,0}$ state.
It is important to emphasize that, because the phonon
modes are {\em collective} oscillations, when the m-th ion
acquires an amplitude in the $\ket{\mbox{1 phonon}}$ state, then
all of the other ions in the trap also acquire an
amplitude in the $\ket{\mbox{1 phonon}}$ state.
This amplitude is dependent on the state of the
m-th ion.  Thus by means of interaction with
the CM phonon mode, the internal state of any ion in the trap
can be changed conditionally on the internal state of any other ion.
The two-level system 
$\left\{ \ket{\mbox{0 phonons}},\ket{\mbox{1 phonon}}\right\}$ 
can be considered as an additional qubit that acts as a quantum bus
channel, that can be used to transfer quantum information between different
ions in the quantum register. 

From these basic building blocks, a CNOT gate
between ions c (``control'') and t (``target'') 
can be constructed from the sequence of five laser
pulses
\be
CNOT_{ct}=
\hat{V}_{t}(\pi/2,\pi/2)
\hat{U}_{c}\left(\pi,0\right) 
\hat{U}^{aux}_{t}\left(2\pi,0\right)
\hat{U}_{c}\left(\pi,0\right)
\hat{V}_{t}(\pi/2,\pi/2).
\ee

\subsection{Readout}
Once the computational state manipulations required are completed, the result must be
read out, which means that the state of each qubit must be measured.
This can be accomplished using the
quantum jump technique, which is relatively straightforward
when using the single laser qubit scheme.  
For example, with $Ca^{+}$ ions, 
each qubit consists of a sub-level of the $4\,^{2}S_{1/2}$ 
ground state, $\ket{0}$, and a sub-level of the $3\,^{2}D_{5/2}$
metastable excited state, $\ket{1}$, (see fig.2).
This qubit can be interrogated with the laser tuned to the $4\,^{2}S_{1/2}$ 
to $4\,^{2}P_{1/2}$ dipole transition at 397nm. 
If the ion radiates, its state has been collapsed to the 
$\ket{0}$ state, 
whereas if it remains dark then its state
has collapsed to the $\ket{1}$ state. The  $4\,^{2}P_{1/2}$ level can also decay
to the  $3\,^{2}D_{3/2}$ level, and so a pump-out laser at
866nm will also be required in order to stop population being trapped 
in that metastable level.

With the Raman scheme, readout is a little more complicated.  Assume that 
$\ket{0}$ and $\ket{1}$ are the $M_{J}=-1/2$ and $M_{J}=+1/2$ 
sublevels of the $4\,^{2}S_{1/2}$ ground state, respectively.
The most straightforward method of performing the readout
in these circumstances is to apply a $\pi$ pulse from the $\ket{1}$ 
(or $\ket{0}$)  state to
a sub-level of the $3\,^{2}D_{5/2}$ state, and then use a 
laser at 397nm to observe fluorescence (if any) due to population in 
the $\ket{0}$ (or $\ket{1}$) state, as in the single laser scheme.
Alternatively, 
a $\sigma^{-}$ circularly polarized laser at 393nm can excite population 
from $\ket{0}$ state to the $M_{J}=-3/2$ sub-level of 
the  $4\,^{2}P_{3/2}$ level, which will decay by dipole emission back to 
$\ket{0}$.  One can avoid exciting population from the $\ket{1}$
level by applying a sufficiently strong magnetic field that the 
transition from this state to 
the  $4\,^{2}P_{3/2}\, M_{J}=-1/2$ sub-level is shifted off resonance
by the Zeeman effect (a magnetic field of 200 Gauss should be sufficient
for this purpose). However, the $4\,^{2}P_{3/2} M_{J}=-3/2$ sub-level can also 
decay to both the $3\,^{2}D_{3/2}$ and $3\,^{2}D_{5/2}$ levels,  
and pumping out the population trapped in these states might be
lead to difficulties.  

\subsection{Tolerances and Laser Requirements}
\subsubsection {Pulse Durations and Standing Waves}
In deriving the formulas that describe the effects of laser pulses
on the qubits we have made various assumptions which
allow us to discard off-resonant terms in the Hamiltonian.
The criteria for making these approximations place constraints
on the experimental parameters, in particular the duration of
the various laser pulses required to perform operations.

For the V-type pulses, we have neglected terms involving 
coupling to 
phonon modes.  These can be neglected provided that 
$(\Omega_{1}/\omega_{x})^{2}\ll 1$.  Using eq. \ref{Omega1},
and the fact that the duration of a V-type $\pi$ pulse is 
$t_{V}=\pi/\Omega_{0}$, we find that the pulse duration
must obey the following inequality:
\be
t_{V}\gg \frac{\pi \eta}{ \sqrt{N}\omega_{x}}.
\label{TottenhamHotspur}
\ee
For 10 $Ca^{+}$ ions in a trap with a $2\pi\times 500$ kHz 
axial angular frequency, eq.(\ref{TottenhamHotspur}) 
implies that  $t_{V}\gg 7.5$ nsec 
for the single laser
scheme and $t_{V}\gg 14$ nsec for the Raman scheme.
This condition will become easier to satisfy as the number of
ions $N$ grows large.

For U-type pulses, we have neglected the terms in the 
interaction Hamiltonian (\ref{Hint2}) which give rise to
``direct'' transitions, i.e. those that do not involve the
creation or annihilation of a phonon.  Using a similar argument,
one can show that the duration of a U-type $\pi$-pulse
must obey the following inequality:
\be
t_{U}\gg \frac{\pi \sqrt{N} }{ \eta\omega_{x}} \;\;\mbox{(traveling 
wave)}.
\label{twlim}
\ee
Thus if we have a large number of ions in our trap, the duration
of the sideband detuned U pulses must become very long.  Using
the same example that we quoted for the V pulses, i.e.
10 $Ca^{+}$ ions in a $2\pi\times 500$ kHz 
trap, (\ref{twlim}) implies that  $t_{U}\gg 130\,\mu$sec
for the single laser
scheme and $t_{U}\gg 72\,\mu$sec for the Raman scheme.

A method by which these pulse durations can be made shorter
 has been suggested by Cirac and Zoller and 
co-workers (\cite{CZ, CiracBlattZollerPhillips}).  If one
were to apply the laser field in a configuration such that $\Omega_{0}$
were zero, but $\Omega_{1}$ was non-zero, then the direct transition
terms in the Hamiltonian (\ref{Hint2}) would be zero.  For example,
the single laser could be used in a standing wave configuration
such that one has a node at the location of the ion one is
trying to address \footnote {Note that the roles of
nodes and anti-nodes are reversed for quadrupole allowed
transitions.  For details, see ref.\cite{dfvj}.}
.  In these circumstances, one would still have to
worry about the possibility of exciting the ``wrong'' phonon
modes, and it can be shown \cite{dfvj} that this places the following
 constraint
on the duration of U-type $\pi$ pulses:
\be
t_{U}\gg \frac{2.6 \pi }{\omega_{x}} \;\;\mbox{(standing
wave)}.
\label{swlim}
\ee
This has the advantage that the duration of U-type pulses
is independent of the number of ions in the trap. 
Again using the example of 10 $Ca^{+}$ ions in a $2\pi\times 500$ kHz 
trap, the durations of the U-pulses must be  $t_{U}\gg 2.6\,\mu$sec
when we have standing waves.
However there are considerable technical difficulties in
arranging a laser beam, which
must have a component of its wavevector parallel to the
trap axis, in a 
standing wave configuration, with all of the ions either
at a node or an anti-node.  Thus for the small number of
ions which will be involved in the first generation
ion-trap quantum computers, it seems that the 
practical problems associated with building an optical system
that allows us to address the ions with standing waves will
outweigh the advantages of having shorter pulses. However, they will probably
be {\em required} if we are to have a computer with more
than a few qubits.

\subsubsection {Laser Power Requirements}
The expressions for the Rabi frequencies, eqs.(\ref{Omega0})
and (\ref{Omega1}), relate the
rate of flipping between the two levels $\ket{0}$ and $\ket{1}$
to the electric field strength of the lasers.  From
these formulas, we can derive an expression relating the 
laser power to the duration of the various pulses.  This
is important information to know when determining the 
specifications of the laser system which must be built.
The power in a Gaussian laser beam is given by
(ref. \cite{MilonniEberly}, eq.14.5.27, p.488)
\be
P=\frac{c\epsilon_{0}}{4}\pi w_{0}^{2}|E|^{2},
\label{power}
\ee
where $c$ is the velocity of light, $\epsilon_{0}$ is
the permittivity of the vacuum and $w_{0}$ is the $1/e^{2}$ radius
of the focal spot.  On substituting from eq.
(\ref{Omega1}), we find the laser power is given by the
following expression:
\be
P\approx\left\{
\begin{array}{ll}
\normalsize\frac{\D
\hbar w_{0}^{2}\omega_{L}\omega_{x}NM}{
\D At^{2}_{U}}&\mbox{(single 
laser)}\\
&\\
\normalsize\frac{\D
\hbar w_{0}^{2}\omega_{L}^{2}\delta NM}{\D c A t_{U}}
\sqrt{NM\hbar\omega_{x}}&\mbox{(Raman)}
\end{array}
\right. .
\ee
We have assumed that, in the Raman case, the pump and
Stokes lasers have approximately the same power. In these
formulas, $t_{U}$ is the duration of a U-type $\pi$-pulse,
i.e. $ t_{U}=\pi/\Omega_{1}$; the other symbols have the
same meaning as given above (see the paragraph following 
eq.(\ref{Omega0})). For 10 Calcium ions, assuming 
a pulse duration $t_{U}$ of 5 $\mu$sec (which is compatible
with the minimum pulse duration when standing waves
are employed), a laser spot size
$w_{0}$ of 10 $\mu$m, an axial frequency $\omega_{x}$ of
$2\pi\times$500 kHz, and a Raman detuning $\delta$ of 
$2\pi\times$100 MHz, the power required for the single laser scheme is
25 mW, compared to a required power of 0.11 $\mu$W for the Raman
scheme.  Longer pulses, which would be required if
standing waves are not used, would need {\em less} power.

\subsubsection{Error Rates and Fault Tolerant Quantum Computing}
In the two years since Cirac and Zoller's original
proposal was published, there have been several calculations of the
limitations of the capabilities of such devices due to various
decoherence effects \cite{HJKLP, PK, HR}.  While
these investigations are useful for identifying theoretically the
various important decoherence processes, none of them take into account
the recent breakthroughs in the realm of fault tolerant quantum
computation \cite {MannyRay, Preskill}.  As mentioned in the 
introduction, it has been demonstrated that, provided quantum gates
can be performed within a certain threshold degree of accuracy (the most 
optimistic estimate of which is $\sim 10^{-6}$), then in principle 
arbitrarily large quantum computations can be performed accurately.
This sets an obvious figure of merit for quantum computation 
technology, namely, the expected probability of error
in one quantum gate.

For trapped ion quantum computers using the single laser addressing
scheme, there are two principal fundamental causes of error which
we must examine: spontaneous emission from the upper level of
the qubit, and breakdown of the two-level system.  The first suggests
that we should use the {\em shortest} possible laser pulses, so
that the gate operation is completed quickly, before the upper level 
can decay.  The second type of error is reduced by using very low laser 
powers, and so {\em long} duration laser pulses should be the best.
Thus there is an optimum situation when these two effects balance.
If we then use the limiting cases given by eqs.(\ref{twlim}) and 
(\ref{swlim}), we obtain the following expression for the minimum
possible error probability per CNOT using $Ca^{+}$ ions:
\be
\varepsilon \sim \left\{\begin{array}{ll}
8.9\times 10^{-6} N^{1/3}&\mbox{(standing wave)}\\
3.6\times 10^{-5} N^{1/2}&\mbox{(traveling wave)}
\end{array}
\right.
\ee
Thus although the use of standing waves will reduce the
error significantly, it still does not appear that it will be
possible to meet the existing accuracy thresholds using single laser
addressing.  (Other species of ion give similar results.)
This, however, does not imply that 
quantum computation experiments using
single laser addressing are without merit; 
discounting the possibility that further 
theoretical advances may reduce the error threshold, 
the single laser system will still be useful as
a experimental device for performing limited
gate operations on a dozen or so qubits, which
will allow one to confirm both the practicality
of quantum computation with cold trapped ions,
as well as aspects of the theory of fault tolerant
quantum computation.  

Using the Raman system,
the fundamental source of error we have to worry
about is spontaneous emission from the upper level $\ket{2}$
of the three level system (see fig. 3.b).  As the average population
in this level and the rate at which the quantum gate
is performed are both proportional to the laser power,
the error rate can be shown to be {\em independent}
of the duration of the laser pulses \cite{MBP}. One can
minimize this error by using the largest possible Raman
detuning $\delta$ and the smallest possible axial frequency
$\omega_{x}$.  However, when one is detuned from one
level by a large amount, one can come into resonance with
another level; hence, when calculating the error, {\em all} of
the ion's levels must be taken into account.  The smallest
practical trap frequency can be estimated to be 100 Hz,
although the experimental problems associated with cooling
at this frequency are more than a little daunting.  In
this case, the error per gate for $Ca^{+}$ ions is
\be
\varepsilon \sim 1.3\times 10^{-8}N^{1/2}.
\ee
This implies that in theory it may be possible to perform
reliable quantum computations, with errors
per gate of the order of $10^{-6}$, with 
$\sim$100 $Ca^{+}$ ions.  Similar calculations give even 
larger numbers ($\sim$5000) for $Be^{+}$ ions.
The estimates should be
viewed with a great deal of caution: throughout the work
done on fault tolerant techniques, the assumption
was made that the error rate per gate was independent of the
number of qubits, and so large numbers of qubits could
be used without undue penalty.  When the error rate
is dependent on the number of qubits, as is the case
here, the results are no longer valid.  However, they
do suggest that some optimism regarding the future of 
ion trap quantum computation may not be unwarranted.

\section{Experimental Considerations}\setcounter{equation}{0}
\subsection {Choice of Ion}

There are three requirements which the species of ion
chosen for the qubits of an ion trap quantum computer
must satisfy:

1. If we use the single laser scheme, the ions must have a level 
that is sufficiently long-lived to allow
some computation to take place; this level can also be used for
sideband cooling.

2. the ions must have a suitable dipole-allowed transition for Doppler
cooling, quantum jump readout and for Raman transitions (if we chose
to use two sub-levels of the ground state to form the qubit);

3.  These transitions must be at wavelengths compatible with current
laser technology.

Various ions used in atomic frequency standards work
satisfy requirement 1: A long-lived transition
will have a very narrow spectral line which
can be exploited for that application.  Examples are
$Hg^{+}$, $Ca^{+}$ or $Ba^{+}$ (which have
quadrupole allowed transitions in the optical, near infrared
or near ultraviolet regions of the spectrum, with lifetimes
ranging from hundreds of milliseconds to several seconds;
for details, see \cite{dfvj}).
An even more exotic possibility is $Yb^{+}$ which has an optical
electric octupole transition with a lifetime of about 10 years 
\cite{Klein}. 
Of these ions, $Ca^{+}$ offers the advantages of transitions
that can be accessed with titanium-sapphire or diode lasers and a reasonably
long-lived metastable state. 
The relevant energy levels of the $A=40$ isotope are shown in fig.2.  
The dipole-allowed transition from the $4\,^{2}S_{1/2}$ ground state to the 
$4\,^{2}P_{1/2}$
level with a wavelength of 397 nm can be used for Doppler cooling and
quantum jump readout; The 732 nm electric quadrupole transition from the 
$4\,^{2}S_{1/2}$ ground state to the $3\,^{2}D_{3/2}$ metastable level 
(lifetime $\approx 1.08 sec.$) is
suitable for sideband cooling. 
In the single laser computation scheme,
the qubits and auxiliary level can be chosen
as the electronic states
\be
\ket{0} =\ket{4\,^{2}S_{1/2},\,M_{j}=1/2}, \: 
\ket{1} =\ket{3\,^{2}D_{5/2},\,M_{j}=3/2}, \: 
\ket{aux}=\ket{ 3\,^{2}D_{5/2},\,M_{j}=-1/2}
\ee

The $\hat{V}$ and $\hat{U}$ operations can be driven using left-handed 
circular
polarized 729-nm radiation, while the $\hat{V}^{aux}$ and
$\hat{U}^{aux}$ operations require
right-handed circular polarization.

This ion can also be used for Raman type qubits, with the two
Zeeman sublevels of the $4\,^{2}S_{1/2}$ ground state forming the
two qubit states $\ket{0}$ and $\ket{1}$, 
with one of the sublevels of the $4\,^{2}P_{1/2}$ level
being the upper level $\ket{2}$.  As mentioned above, a magnetic field of
200 Gauss should be sufficient to split these two levels
so that they can be resolved by the lasers.  The pump and Stokes beams 
would be formed by splitting a 397nm laser into two, and
shifting the frequency of one with respect to the other
by means of an acousto-optic or electro-optic modulator.  This arrangement
has a great advantage in that any fluctuations in the
phase of the original 397nm laser will be passed on to
both the pump and Stokes beams, and will therefore be
canceled out, because the dynamics is only sensitive
to the difference between the pump and Stokes phases.
One problem in realizing the Raman scheme in $Ca^{+}$
is the absence of a third level in the ground state that
can act as the auxiliary state $\ket{aux}$ required for
execution of quantum gates.  This difficulty could be removed
by using the alternative scheme for quantum logic recently
proposed by Monroe {\it et al.} \cite{NISTsimp}; alternatively, one could 
use an isotope of $Ca^{+}$ which has non-zero nuclear spin, thereby
giving several more sublevels in the ground state due to the hyperfine
interaction \footnote{The only such stable isotope of Calcium is
$^{43}Ca$, which has nuclear spin 7/2 and a natural abundance of 0.135$\%$.
Enriched samples are available at a cost of $\$$441 per mg 
\cite{Morgan}.}; other possibilities that have been suggested
for an auxiliary state with $^{40}Ca^{+}$ in the Raman scheme
are to use a state of a phonon mode other than the CM mode 
\cite{Steaneaux}
or one of the sublevels of the $3\,^{2}D$ doublet \cite{Blattaux}.

\subsection {The Radio Frequency Ion Trap}
Radio-frequency (RF) quadrupole traps, also named ``Paul traps'' after their
inventor, have been used for many years to confine
electrically charged particles \cite{Paul} (for an introduction
to the theory of ion traps, see refs. \cite{NISTtrap, Ghosh}).
The classic design of such a Paul trap
has a ring electrode with endcap electrodes above and below, with
the ions confined to the enclosed volume. A single ion can be located
precisely at the center of the trap where the amplitude of the RF field is
zero. But when several ions are placed into this trapping field, their
Coulomb repulsion forces them apart and into regions where they are
subjected to heating by the RF field. For this reason in our experiment
ions are confined in a linear RF quadrupole trap.
Radial confinement is achieved by a quadrupole RF
field provided by four 1 mm diameter rods in a rectangular
arrangement.  Axial confinement is provided by DC voltages applied
to conical endcaps at either end of the RF structure; the endcap
separation is 10 mm.
The design of the trap used in these experiments is shown in
diagrammatically in Figs.4 and 5.  An image of a trapped ion is
shown in Fig.6.

The main concerns for the design are to provide sufficient
radial confinement to assure that the ions form a string on the
trap axis after Doppler cooling; 
to minimize the coupling between the radial and axial
degrees of freedom by producing radial oscillation frequencies
significantly greater than the axial oscillation frequencies;
to produce high enough axial frequencies to allow the use of
sideband cooling\cite{NISTcool89}; and to provide sufficient spatial
separation
to allow individual ions to be addressed with laser beams.

\subsubsection {Radial Confinement}

A sinusoidal voltage with a DC offset, $\Phi=\Phi_{DC}-\Phi_{RF}\cos
(\omega_{RF} t)$,
is applied to two opposite rods of the RF structure while the other
two are grounded.  The potential near the center of the trap is
\begin{equation}
\Phi = \Phi_o \frac{z^2-y^2}{2r_o^2},
\label{eq:pot}
\end{equation}
\noindent where $r_o$ is a constant dependent on
the distance from the trap axis to the RF rods
and $y$ and $z$ are the distances from the trap axis along the y and 
z axes, respectively (see fig.1).
For this trap, $r_o=$1.4mm.

The equation of motion for an ion in this time-dependent quadrupole
potential is the well known Matthieu's equation, the solutions of
which \cite{AbSt} are parameterized in this case, by the quantities
$a=4e\Phi_{DC}/(M \omega_{RF}^2 r_o^2)$ and
$q=2e\Phi_{RF}/(M \omega_{RF}^2 r_o^2)$.  For
a proper choice of these parameters, the motion of the ion can be
treated by a pseudopotential approach.  Here, the ions are described
as moving in the effective quadratic potential
\begin{equation}
\Phi^{(eff)}=\frac{e\Phi_{RF}^2\rho^2}{4M\omega_{RF}^2 r_o^4},
\label{eq:pseudo}
\end{equation}
where $\rho=\sqrt{y^{2}+z^{2}}$ is the radial distance from the trap
axis.  The total motion consists of a
fast ripple superimposed on a slower, larger amplitude oscillation,
referred to as the secular motion, of frequency
\begin{equation}
\omega_{r}=\frac{e\Phi_{RF}}{\sqrt{2}M\omega_{RF}r_o^2}.
\label{eq:sec}
\end{equation}

Typical operating parameters for the trap described are $\Phi_{DC}=0$,
$\Phi_{RF}=500$ Volts, and $\omega_{RF}=(2\pi)\times$ 11.5 MHz.  This yields a
radial
pseudowell of 15 eV depth and a secular motion frequency of
$\omega_{r} = (2\pi)\times$ 5 MHz.
Laser cooling of the ions to the Doppler limit imposed by the
natural linewidth of the $4\,^{2}S_{1/2}-4\,^{2}P_{1/2}$ cooling
transition will
yield a temperature $T_{Dop} \approx 85$ $\mu$K.  Comparing this
thermal energy with the radial pseudowell depth, we find that the
ions should be confined to within roughly 30 nm of the trap axis
after Doppler cooling.

\subsubsection {Axial Confinement}

As mentioned above, axial confinement is provided by conical electrodes
at either end of the RF structure.  The conical shape allows maximum
laser access while providing a large region at the trap center for
which the potential is harmonic.  A static bias voltage of up to 500 V
is applied to both endcaps.

When the ions are sufficiently cold, they will crystallize to form
a line or ``string of pearls''.  In this configuration
the axial force on the m-th ion due to the harmonic binding potential
and the Coulomb force upon it due to all of the other ions is given
by the formula
\be
F_{m}=M\omega_{x}^{2}x_m+
\sum_{\stackrel{\scriptstyle n=1}{m \neq n}}^N
\frac{e^2}{4\pi\epsilon_0}\frac{1}{(x_n-x_m)^{2}} \ ,
\ee
where the symbols have the same meaning as in eq.(\ref{potty}).  
The equilibrium positions of
the ions are determined by  the set of $N$ equations, $F_{m}=0,\;
(m=1,2,\ldots N)$.  For $N=2$ or $N=3$, the equations may be
solved analytically; for larger values of $N$ numerical solutions
must be found.  For details, see ref. \cite{dfvj}.
We find that the minimum inter-ion spacing (which will
occur at the middle of the string) for
$N$ ions and axial frequency
$\omega_{x}$ is
given by:
\be
x_{min}(L)=\left(\frac{e^2}
{4\pi\epsilon_0 M\omega_{x}^2}\right)^{1/3}
\frac{2.018}{N^{0.559}} \ . \\
\ee
This relationship is  important in determining the
extent of ``cross-talk'' error in a quantum
computer, due to the focal
region of lasers overlapping more than one ion.
 \cite{HJKLP}.
Figure 7 shows this calculated minimum equilibrium separation
as a function of axial oscillation
frequency, which is proportional to the square root of the applied
trapping voltage. The
curve is truncated at $(2\pi)\times$500 kHz, because, 
as discussed earlier we are
interested only in frequencies significantly lower than the 5 MHz
secular frequency.
Preliminary numerical model 
calculations \cite{BlattNaegerl} for our trap
geometry indicated
that an endcap voltage of 150 V would yield an axial trapping frequency
of roughly $(2\pi)\times$200 kHz.  
We have experimentally verified this prediction by observing resonance ion
heating through parametric excitation of the axial oscillation of the ions.
Small number of ions were trapped and laser cooled. A weak RF drive was
coupled to the DC input applied to the end caps. The frequency was slowly
varied and when the applied frequency was equal to $2 \times \omega_x$ a
strong heating of the ion cloud was observed. Figure 8 shows the results of
these tests in comparison to the numerical calculations. The solid line is
a fit to the data
assuming a strict quadratic relation between axial frequency and trap
potential.
The proportionality constant gives the shielding of the external potential
by the trap electrodes.
As can be seen, when there are two ions in the trap
a frequency of $(2\pi)\times$200 kHz 
corresponds to an ion separation of
approximately 20 $\mu$m.  Hence, to resolve the axial motion
sidebands on the ion fluorescence signal and to cool the
ions to the vibrational  ground level of the axial well, laser linewidths
well under $(2\pi)\times$200 kHz will be required.  Furthermore, gate operations
involving manipulations of individual ions will require laser spot
sizes well under 20 $\mu$m.

The linear configuration of the ions will break down
if there are too many ions in the trap.  The Coulomb force
from ions at the ends of the string will become so
great that the radial confinement will become unstable,
and the ions will adopt a zig-zag configuration, which is
susceptible to RF heating.  The radial confinement
is characterized by a harmonic frequency $\omega_{r}$,
given by eq. (\ref{eq:sec}).  Numerical modeling of the
zig-zag breakdown (based on the instability of the transverse
oscillation modes) gives the following formula for the maximum
number of ions that can reside in a linear configuration:
\be
N_{max}=1.82 \left(\frac{\omega_{r}}{\omega_{x}}\right)^{1.13}.\
\ee
This formula is in general agreement with that previously  worked out by
numerical modeling of cold confined plasmas \cite{Schiffer}.
In the ion trap used in our
experiment, $\omega_{r}=2\pi\times 5 {\rm MHz}$ and
$\omega_{x}$ can be in the range from $2\pi\times 500 {\rm kHz}$ to
$2\pi\times 100 {\rm kHz}$.  Thus the largest number of ions that
can be stored in this specific system in a linear configuration varies
between 24 and 151 ions
respectively.

\subsubsection {Thermalization of Trapped Ions and Noise Driven
Decoherence}

We have theoretically investigated the effects of
external noise sources (as opposed to trap RF effects)
in regard to ion heating and decoherence \cite{Lamore}.
Of the sources considered, which included mechanical vibrations,
black body radiation, and Johnson noise, only the
latter was found to be significant.  By use of
a result based on a classical model of the trapped
ion \cite{WineDehm}, Wineland {\em et al.} \cite{WineBoll} concluded
that Johnson noise cannot account for the observed
ion heating rate.  We have carried the calculation
further and have now shown that the heating rate
(which in the case of Johnson noise is roughly
equal to the harmonic oscillator superposition
state decoherence rate) can be explained within
a reasonable range of system parameters.  

The heating rate is proportional to the effective
resistances connected across the trap electrodes.
However, the fact that this resistance depends 
on the frequency of the noise fluctuations was
not taken into account in \cite{WineBoll}
where it was concluded that the observed heating
rate in a particular experiment was roughly 10 to 100
times faster than predicted by the use of the model in
\cite{WineDehm}.  Our primary conclusions are that
it is incorrect
to use the DC resistance of the trap electrode interconnections
to determine the heating rate, and that in many instances,
better insight  can be obtained through consideration of
the noise correlation time.  We hope to experimentally
investigate the implications of our theoretical work
in the near future.

\subsection {Laser Systems}

The relevant optical transitions for $Ca^{+}$ ions are shown in 
fig.2.  There are 
four different optical processes employed in the quantum computer; each places 
specific demands on the laser system.

The first stage is to cool a small number of ions to their Doppler limit in the ion 
trap, as shown in fig.(9.a).  This requires a beam at 397 nm, the 
$4\,^{2}S_{1/2} - 4\,^{2}P_{1/2}$
resonant transition.  Tuning the laser to the red of the transition causes the 
ions to be slowed by the optical molasses technique \cite{StenholmCool}.  
In this procedure, a 
laser beam with a frequency slightly less than that of the resonant transition of an 
ion is used to reduce its kinetic energy.  Owing to the Doppler shift of the photon 
frequency, ions preferentially absorb photons that oppose their 
motion, whereas they re-emit photons in all directions, 
resulting in a net reduction in momentum along the direction of the
laser beam.  Having carefully selected the trap 
parameters, many cycles of absorption and re-emission will bring the system to 
the Lamb-Dicke regime, leaving the ions in a string-of-pearls geometry.

In order to Doppler cool the ions, the demands on the power and linewidth of the 
397 nm laser are modest.  The saturation intensity of $Ca^{+}$ ions is 
$\sim 10\; {\rm mW/cm}^{2}$, and the laser linewidth must be less than $\sim 
10\; {\rm MHz}$. 
 An optogalvonic 
signal obtained with a hollow cathode lamp suffices to set the frequency.  We 
use a Titanium:Sapphire (Ti:Sapphire) laser (Coherent CR 899-21) with an internal 
frequency doubling crystal to produce the 397 nm light.

During the Doppler cooling, the ions may decay from the $4\,^{2}P_{1/2}$ state 
to the $3\,^{2}D_{3/2}$
state, whose lifetime is $\sim 1 {\rm sec}.$  
To empty this metastable state, we use a second 
Ti:Sapphire laser at 866 nm.  

Once the string of ions is Doppler cooled to the Lamb-Dicke regime, the second 
stage of optical cooling, sideband cooling, will be used to reduce the collective 
motion of the string of ions to its lowest vibrational level 
\cite{WinelandItano}, illustrated in fig. 
(9.b).  In this regime, a narrow optical transition, such as the 732 nm 
$4\,^{2}S_{1/2} - 3\,^{2}D_{3/2}$ dipole forbidden transition,
develops sidebands above and below the 
central frequency by the vibrational frequencies of the ions.  The sidebands 
closest to the unperturbed frequency correspond to the CM vibrational motion.  
If $\omega_{0}$ is the optical transition frequency and $\omega_x$ 
the frequency of the CM 
vibrational motion,  the phonon number is increased by one, unchanged, or 
decreased by one if an ion absorbs a photon of frequency 
$\omega_{0}+\omega_x$, 
$\omega_{0}$ or $\omega_{0}-\omega_x$, 
respectively.  Thus, sideband cooling is accomplished by optically cooling the 
string of ions with a laser tuned to $\omega_{0}-\omega_x$.  

The need to resolve the sidebands of the transition implies a much more stringent 
requirement for the laser linewidth; it must be well below the CM mode vibrational 
frequency of $\sim  (2\pi)\times 200\; {\rm kHz}$.  The laser power 
must also be greater in order to pump 
the forbidden transition.  We plan to use a Ti:Sapphire 
laser locked to a reference cavity to 
meet the required linewidth and power.  At first glance it would seem that, with 
a metastable level with a lifetime of 1s, no more than 1 phonon per second could 
be removed from a trapped ion.  A second laser at 866 nm is used to couple the 
$4\,^{2}P_{1/2}$ state to the $3\,^{2}D_{3/2}$ state to reduce the effective 
lifetime of the D state and 
allow faster cooling times.  The transitions required for realization 
of quantum logic gates and for readout, discussed in detail in
sections 5.2 and 5.3, are shown in fig.(9.c).  These
can be performed with the same lasers used in the Doppler and sideband
cooling procedures.

There are two other considerations concerning the laser systems for 
quantum computation which should be mentioned.  To reduce the total
complexity of the completed system, we are 
developing diode lasers and a frequency doubling cavity to handle the Doppler 
cooling and quantum jump read out.  Also complex 
quantum computations would require that the laser on the  
$4\,^{2}S_{1/2} - 3\,^{2}P_{5/2}$ 
computation transition have a coherence time as long as the computation time.  
This may necessitate using qubits bridged by Raman transitions as 
discussed above, which 
eliminates the errors caused by the phase drift of the laser. 

\subsection {Qubit Addressing Optics}
	In order for the $Ca^{+}$ ion qubits to be useful for actual calculations, it 
will be necessary to address the ions in a very controlled 
fashion.  
Our optical system for qubit addressing is shown schematically in 
fig. 10.  
There are two aspects to be considered in the
design of such a system: the precise interactions with a 
single ion; and an arrangement for switching between 
different ions in the string.  In addition to the obvious constraints on 
laser frequency and polarization, the primary consideration for making
exact $\pi$- or $2\pi$-pulses is control of the area (over time) of the driving 
light field pulse.  The first step toward this is to stabilize the 
intensity of the laser, as can be done to better than 
$0.1\%$, 
using a standard ``noise-eater".  Such a device typically consists of an 
electro-optical polarization rotator located between two polarizers; the 
output of a fast detector monitoring part of the transmitted beam is used 
in a feedback circuit to adjust the degree of polarization rotation, and 
thus the intensity of the transmitted light.  Switching the light 
beam on and off 
can be performed with a similar (or even the same)
device.  Because such switches can possess rise/fall times on the scale of 
nanoseconds, it should be possible to readily control the area under the 
pulse to within $\sim 0.1\%$, simply by accurately determining the width of the 
pulse.  A more elaborate scheme would involve an integrating detector, 
which would monitor the actual integrated energy under the pulse, shutting 
the pulse off when the desired value is obtained.  

	Once the controls for addressing a single ion are decided, the means for 
switching between ions must be considered.  Any system for achieving this 
must be fast, reproducible, display very precise aiming and low 
``crosstalk" (i.e. overlap of the focal spot onto more than one ion),
and be as simple as possible.  In particular, it is desirable 
to be able to switch between different ions in the string in a time short 
compared to the time required to complete a given $\pi$-pulse on one ion.  
This tends to discount any sort of mechanical scanning system.  
Acousto-optic deflectors, which are often used for similar purposes, may 
be made fast enough, but introduce unwanted frequency shifts on the 
deviated beams.  As a tentative solution, we propose to use an 
electro-optic beam deflector, basically a prism whose index of 
refraction, and consequently whose deflection angle, is varied slightly by 
applying a high voltage across the material; typical switching times for 
these devices is 10 nanoseconds, adequate for our purposes\footnote{  
Note that while one could in principle control the pulse area merely by 
swinging the pulse on and off a given ion, the complications involved with 
exposing the ion to a varying spatial distribution make this 
undesirable.}.  One such device produces a maximum deflection of 
$\pm$9 
mrad, for a $\pm$3000V input.  The associated maximum number of resolvable 
spots (using the Rayleigh criterion) is of order 100, implying that 
$\sim$ 20 
ions could be comfortably resolved with negligible crosstalk.

	After the inter-ion spacing has been determined, i.e., by the trap 
frequencies, the crosstalk specification determines the maximum spot size 
of the addressing beam.  For example, for an ion spacing of 20 $\mu$m, 
any spot size (defined here as the $1/e^2$ diameter) less than 21.6 $\mu$m
will yield a crosstalk of less than 0.1$\%$, assuming a purely Gaussian 
intensity distribution (a good approximation if the light is delivered 
from a single-mode optical fiber, or through an appropriate spatial 
filter).  In practice, scattering and other experimental realities will 
increase this size, so that it is prudent to aim for a somewhat smaller spot 
size, e.g. 10 $\mu$m. One consideration when such small spot sizes are 
required is the effect of lens aberrations, especially since the spot must 
remain small regardless of which ion it is deflected on.  Employing 
standard ray-trace methods, we have found that the blurring effects of 
aberrations can be reduced if a doublet/meniscus lens combination 
(figs.11.a) is used (assuming an input beam size of 3mm, and an 
effective focal length of 
30mm).	A further complication is that, in order to add or 
remove phonons from the 
system, the addressing beams must have a component along the longitudinal 
axis of the trap.  Calculations indicate 
that an angle of only about $10^{o}$ between the pump beam and the normal to 
the ion string is sufficient for adequate coupling to the phonons.  
Nevertheless the addressing optics must accommodate
a tilted line of focus, otherwise the intensity at each ion would be 
markedly different, and the crosstalk for the outermost ions would become 
unacceptable.  According to ray-trace calculations, adding a simple wedge 
(of $\sim 2^{o}$) solves the problem (see Figs.11.b) and this has been
confirmed by measurements using the mock system (Fig.11.c).  

	Depending on the exact level scheme being considered, it may be necessary 
to vary the polarization of the light (e.g., from left- to 
right-circularly polarized).  Because the electro-optic deflector requires 
a specific linear polarization, any polarization-control elements should 
be placed after the deflector (see, for example, Fig. 10).  The final 
result is a highly directional, tightly-focused beam with controllable 
polarization and intensity.  

\subsection{Imaging System}
In order to determine the ions' locations and to readout the result
of the quantum computations, an imaging system is required.
Figure 12 shows our current imaging system, which consists of two lenses, one of
which is mounted inside the vacuum chamber, and a video camera coupled to a
dual-stage micro-channel plate (MCP) image intensifier. The first lens with
focal length 15 mm collects the light emitted from the central trap
region with a solid angle of approximately 0.25 sr. The image is relayed
through a 110mm/f2 commercial camera lens to the front plate of the MCP.  This set-up produces
a magnification of 7.5 at the input of the MCP. The input of the 110 mm lens
is fitted with a 400nm narrow band filter to reduce background from the IR
laser and from light emanating from the hot calcium oven and the electron
gun filament.

The dual plate intensifier is operated at maximum gain for the highest
possible sensitivity. This allows us to read out the camera at normal video
rate of 30 frames ${\rm s}^{-1}$ into a data acquisition computer. Averaging and
integrating of the signal over a given time period can then be undertaken by
software. We find this arrangement extremely useful in enabling us to
observe changes of the cloud size or the intensity of the fluorescence with
changes of external parameters like trapping potential, laser frequency,
laser amplitude, etc. in real time.

The spatial resolution of the system is limited by the active diameter of
individual channels of the MCP of approximately 12 $\mu$m. Since the gain
is run at its maximum value cross talk between adjacent channels in the
transition between the first and second stage is to be expected. This
results in the requirement that two incoming photons can only be resolved
when they are separated at the input of the MCP by at least two channels,
i.e. by 36 $\mu$m in our case. With the magnification of the optical system
of 7.5 this translates into a minimum separation of two ions to be resolved
of 5 $\mu$m, which is well below the separation of ions in the axial well of
about 25 $\mu$m expected in our experiment.

\section{Summary and Conclusions}\setcounter{equation}{0}
It is our contention that currently the ion trap proposal
for realizing a practical quantum computer offers
the best chance of long term success.  This in no way is intended
to trivialize research into the other proposals discussed
in section 4: in any of these schemes technological advances 
may at some stage lead to a breakthrough.
In particular, Nuclear Magnetic Resonance does seem to be 
a relatively straightforward
way in which to achieve systems containing a few qubits. 
However, of the technologies which have so far been used
to demonstrate experimental logic gates, ion traps seem to offer the
least number of technological 
problems for scaling up to 10's or even 100's of qubits.

In this paper we have described in some detail the experiment we
are currently developing to investigate the feasibility of
cold trapped ion quantum computation.  We should emphasize
that our intentions are at the moment exploratory: we have
chosen an ion on the basis of current laser technology, rather
than on the basis of which ion which will give the best performance
for the quantum computer. Other species of ion may well give better
performance:  In particular, as mentioned in section 5,
Beryllium ions do have the
potential for a significantly lower error rate due to spontaneous 
emission, although it is
also true that lighter ions may be more susceptible to heating. 
Other variations, such as the use of Raman transitions in place of
single laser transitions, or the use of standing wave lasers need to 
be investigated. Our choice of Calcium will allow us to 
explore these issues. Furthermore, calculations suggest that it
should be possible to trap 20 or more Calcium ions in a
linear configuration and manipulate their quantum states
by lasers on short enough time scales that many quantum
logic operations may be performed before coherence is lost.
Only by experiment can the theoretical estimates of performance
be confirmed. Until all of the sources of experimental error
in real devices are thoroughly investigated, it will be impossible
to determine what ion and addressing scheme enables one 
to build the best quantum computer 
or, indeed, whether
it is possible to build a useful quantum computer with cold trap ions 
at all.

\section*{Acknowledgments}
The authors would like to thank
James Anglin,
Howard Barnum,
Rainer Blatt,
Mark Bocko,
Gavin Brennen,
Carlton Caves,
Ike Chuang,
Ignacio Cirac,
Ivan Deutsch,
David DiVincenzo,
Artur Ekert,
Heidi Fearn,
Salman Habib,
Timothy Havel,
Richard Jozsa,
Hugh Klein,
Manny Knill,
Marvin Kruger,
Norm Kurnit,
Raymond Laflamme,
Stephen Lea, 
Todd Meyrath,
Chris Monroe,
George Morgan,
Keith Miller,
Peter Milonni,
Michael Neilsen,
Albert Petschek,
Martin Plenio, 
John Preskill,
Bernard Rosen,
Barry Sanders,
Andrew Steane,
David Wineland,
Tanya Zelevinsky,
Peter Zoller,
and Wojciech Zurek
for useful discussions and helpful comments.
This research was funded by the National Security Agency.

\newpage
\section*{Figure Captions}

\noindent
Figure 1. A schematic illustration of an idealized laser-ion interaction 
system; ${\bf k}_{L}$ is the wavevector of the single
addressing laser.

\vspace{10mm}
\noindent
Figure 2. The lowest energy levels of $^{40}Ca^{+}$ ions, 
with transition wavelengths and lifetimes listed.  See
\cite{dfvj} for references to the data.

\vspace{10mm}
\noindent
Figure 3.  A schematic illustration of (a) single laser and
(b) Raman qubit addressing and control techniques.

\vspace{10mm}
\noindent
Figure 4 (a) Side view diagram of the linear RF trap used to confine
$Ca^{+}$ ions in these experiments.  The endcap separation is 10 mm
and the gap between the RF rods is 1.7 mm.
(b) End-on view of the linear RF trap electrodes.

\vspace{10mm}
\noindent
Figure 5 (a) Photograph of the trap assembly
(b) Photograph of the assembled ion trap vacuum system.

\vspace{10mm}
\noindent
Figure 6.  Image of a trapped Calcium ion

\vspace{10mm}
\noindent
Figure 7. Calculated minimum ion separation as a 
function of axial oscillation 
frequency for  $Ca^{+}$ ions in a harmonic well.  The frequency is 
proportional to the applied endcap bias voltage and is dependent
on the trap geometry.

\vspace{10mm}
\noindent
Figure 8. Frequency of the axial motion of an ion cloud 
in the linear RF trap. The
diamonds are the data points measured by resonant excitation which are to
be compared to the results of an early model calculation for our geometry
(solid circles). The line is a fit to a quadratic relation between
frequency and voltage, the proportionality factor gives the effective
strength of the axial potential due to the shielding by the RF rods.

\vspace{10mm}
\noindent
Figure 9.  Different transitions between the levels
of $Ca^{+}$ ions required for (a) Doppler cooling,
(b) Resolved sideband cooling and (c) quantum logic operations
and readout.  The single laser addressing technique
has been assumed.

\vspace{10mm}
\noindent
Figure 10.Illustration of the laser beam control
optics system.

\vspace{10mm}
\noindent
Figure 11. Ray trace diagram of the laser beam control
optics  in the focal region; a) doublet meniscus wedge optics
for ion addressing; b) ray-trace diagram in the focal region, 
showing the ability of the system to address individual ions;
c) normalized profile from an experimental test of addressing optics,
using a 670nm 
diode laser and a 5 $\mu$m diameter scanning pinhole.  In this case
the maximum cross talk was 0.34\%.  

\vspace{10mm}
\noindent
Figure 12. Illustration of the ion imaging system.

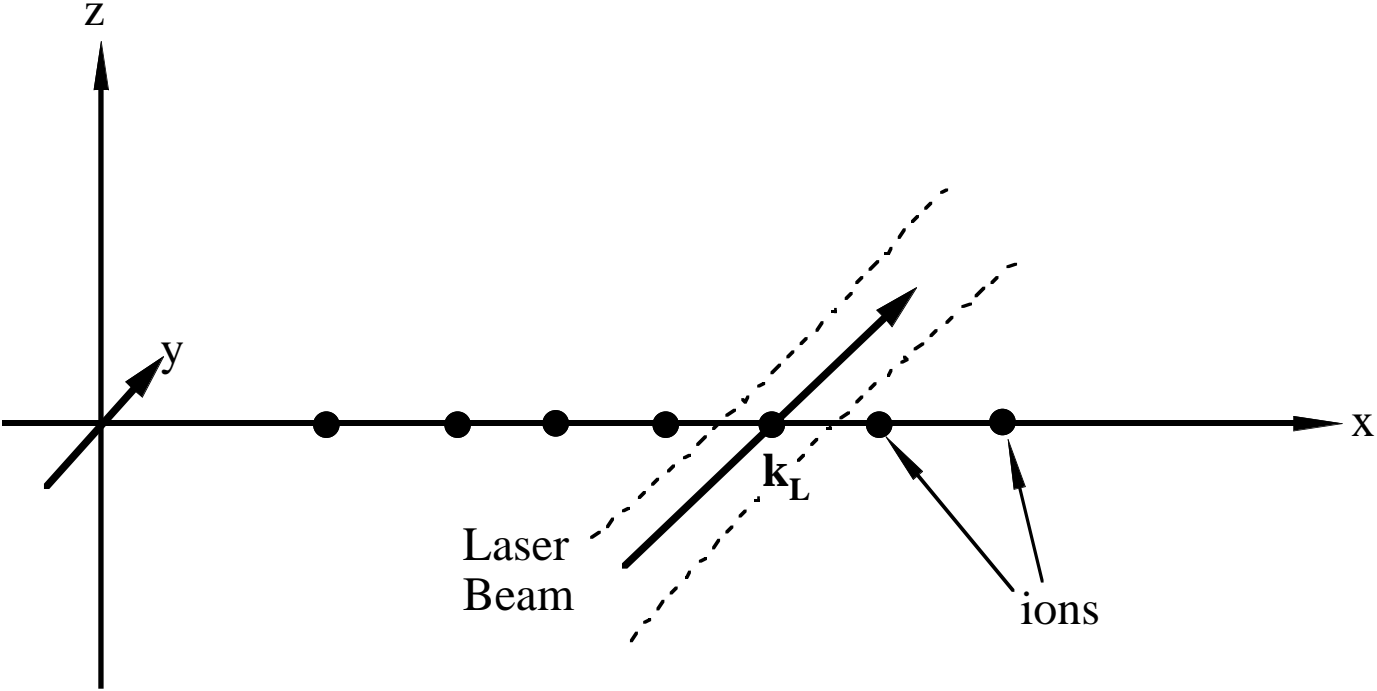
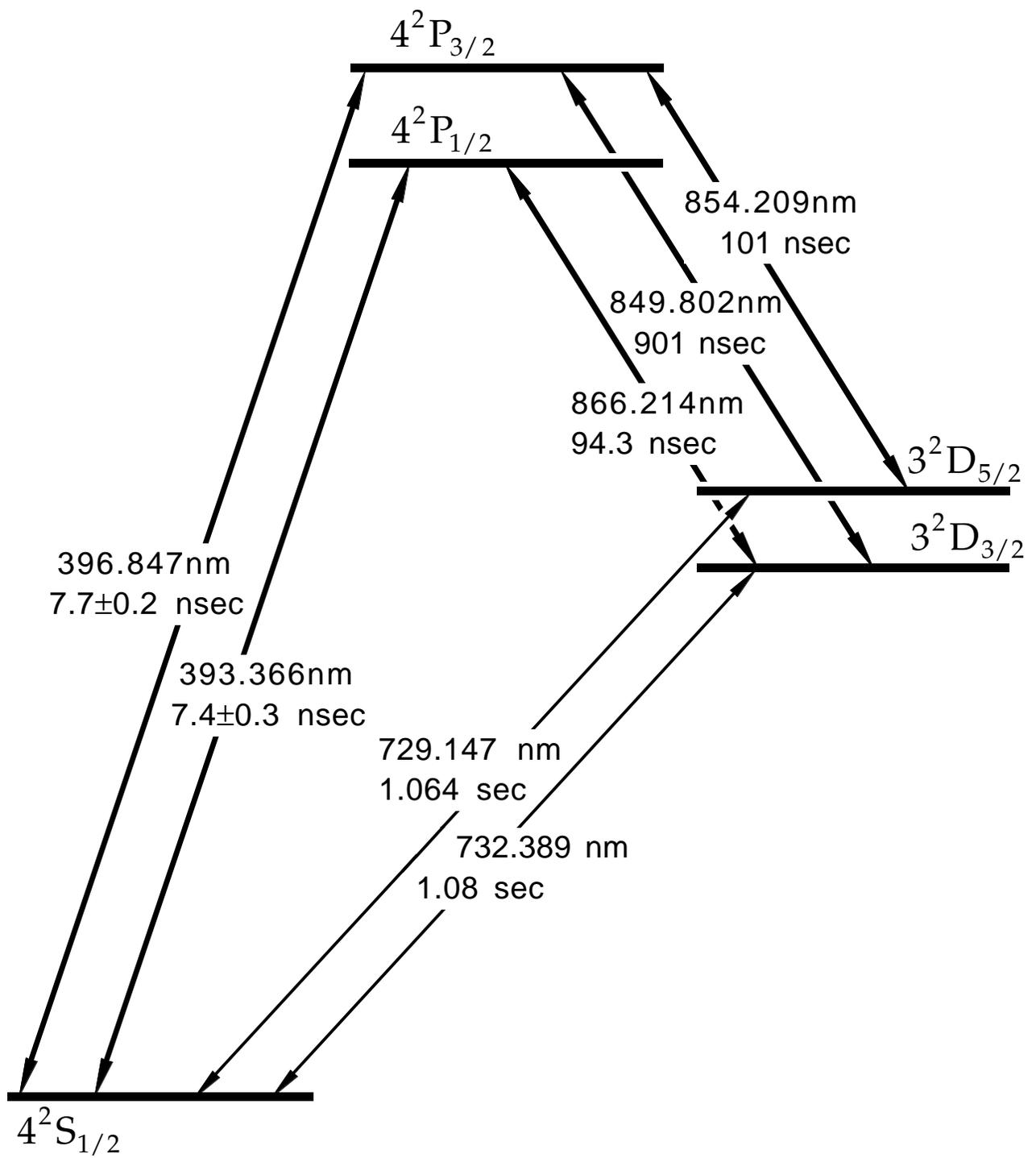
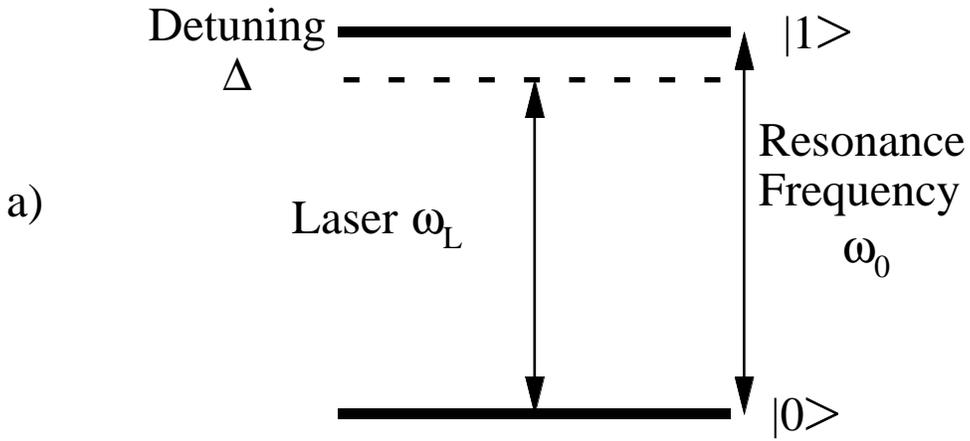
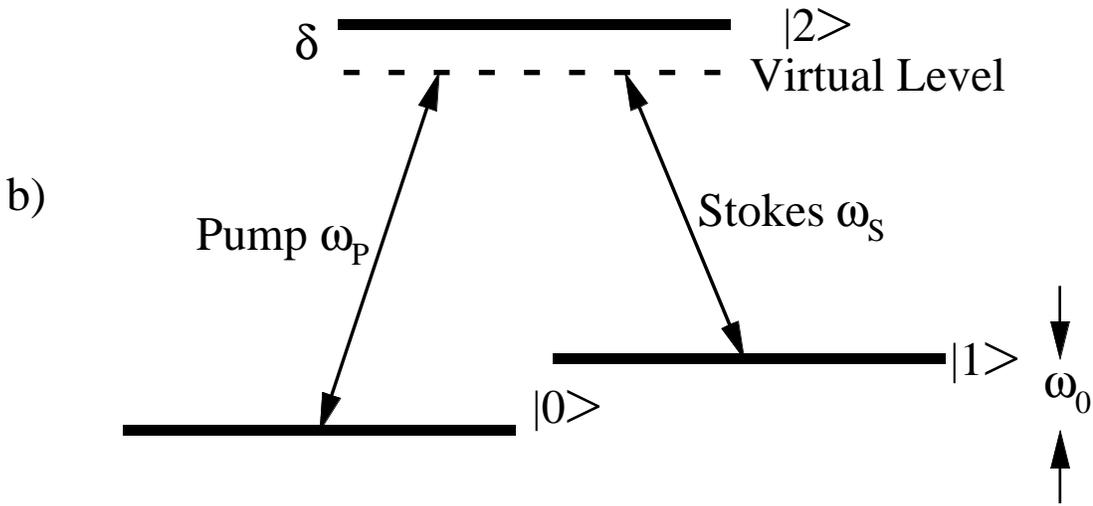
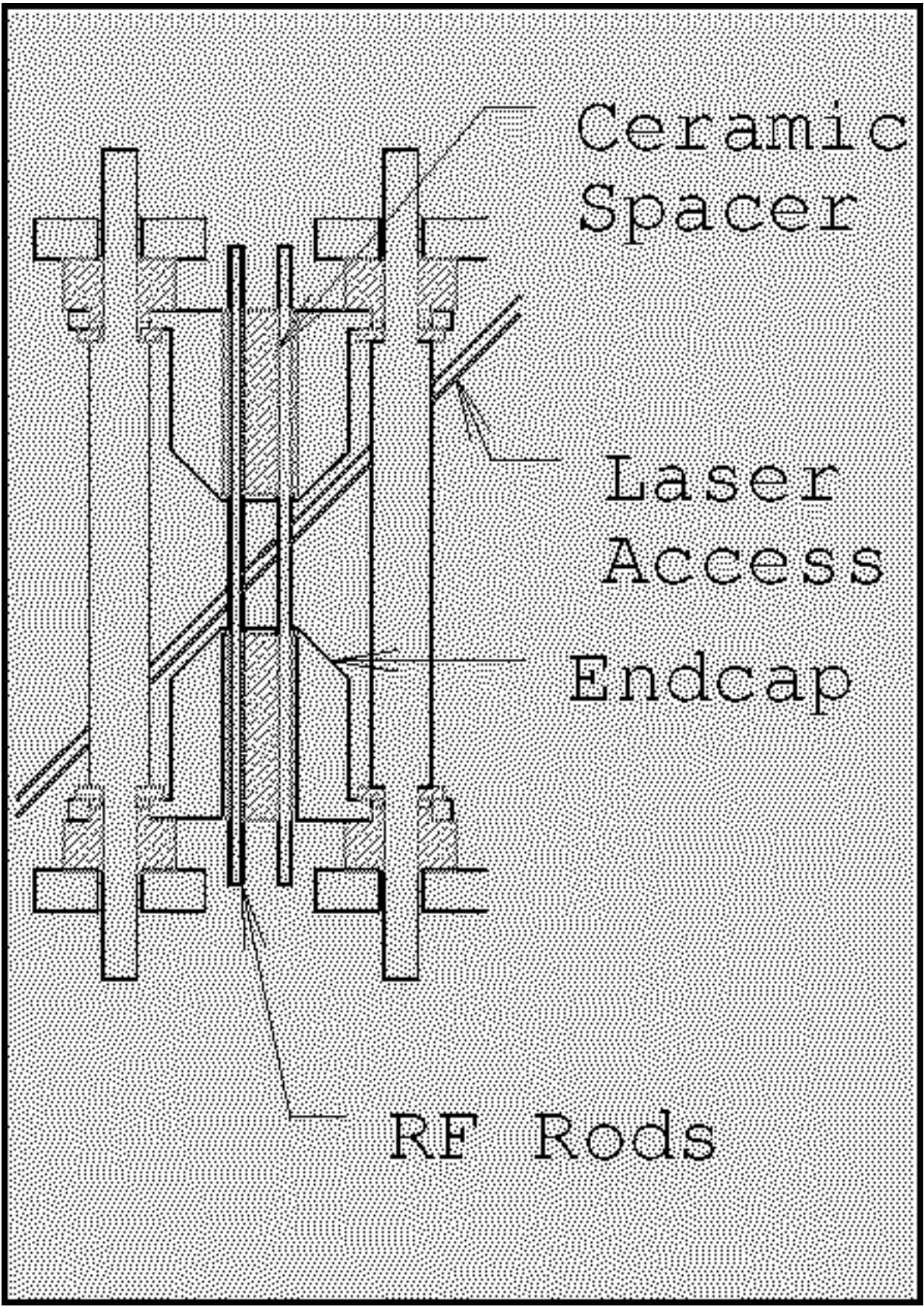
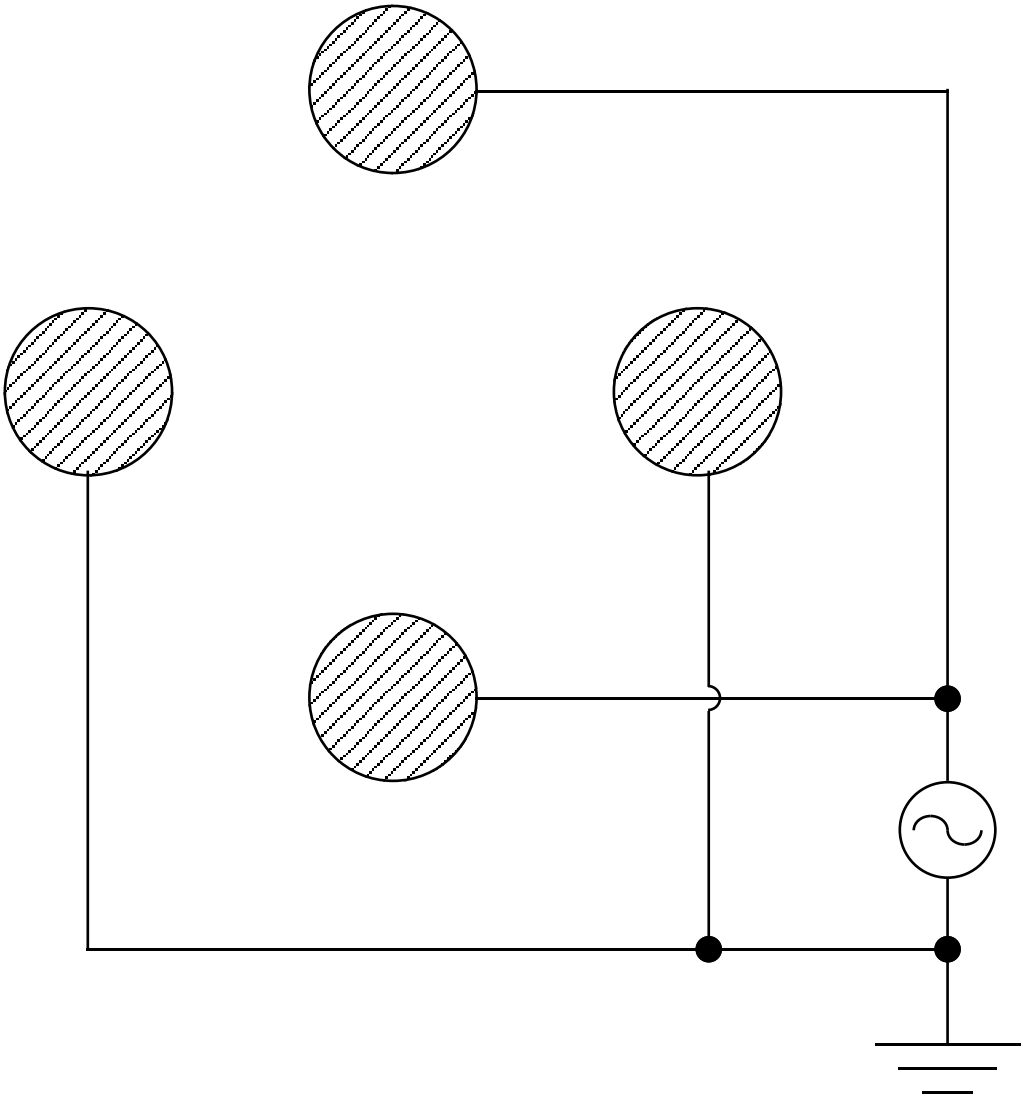
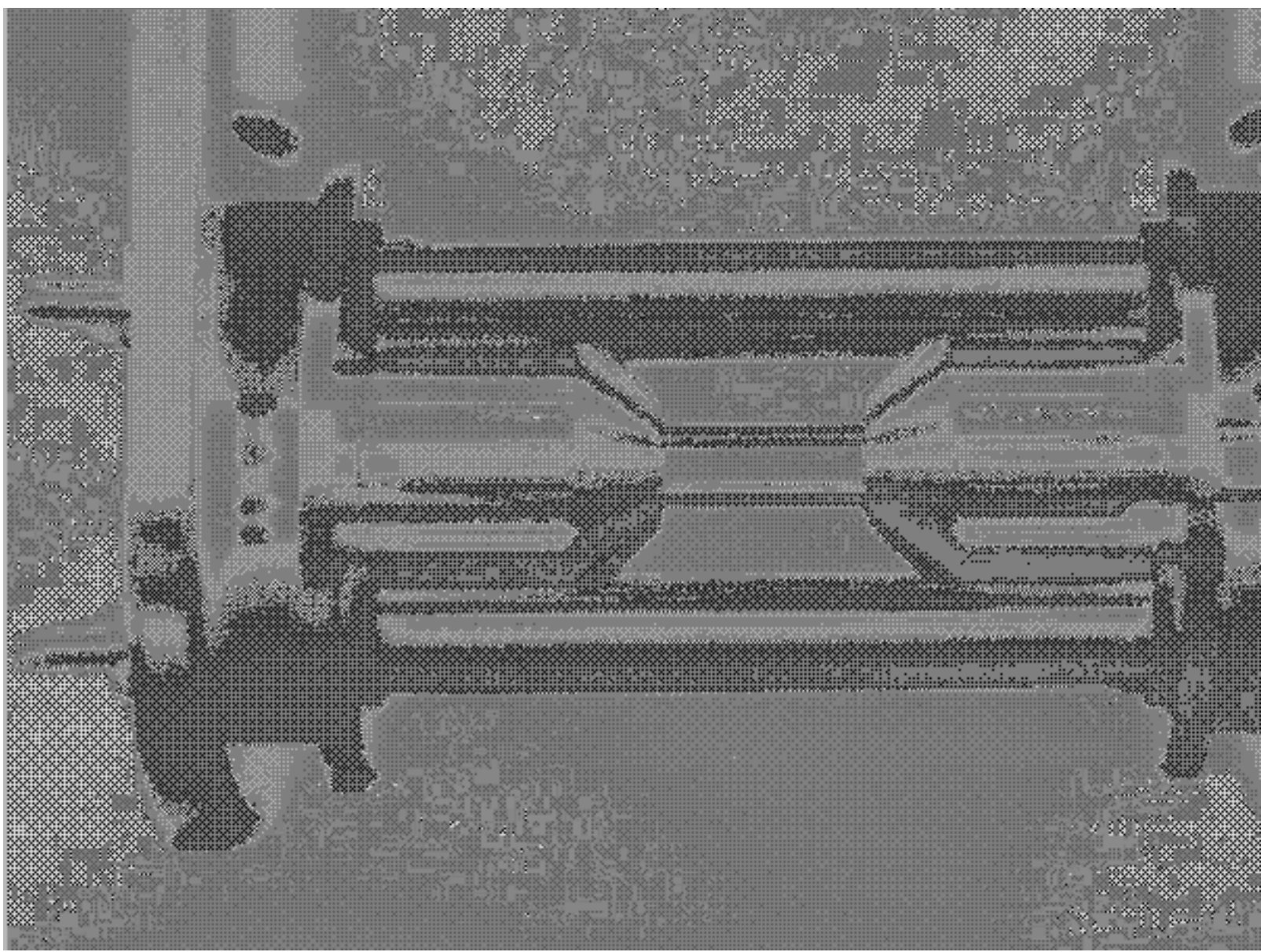
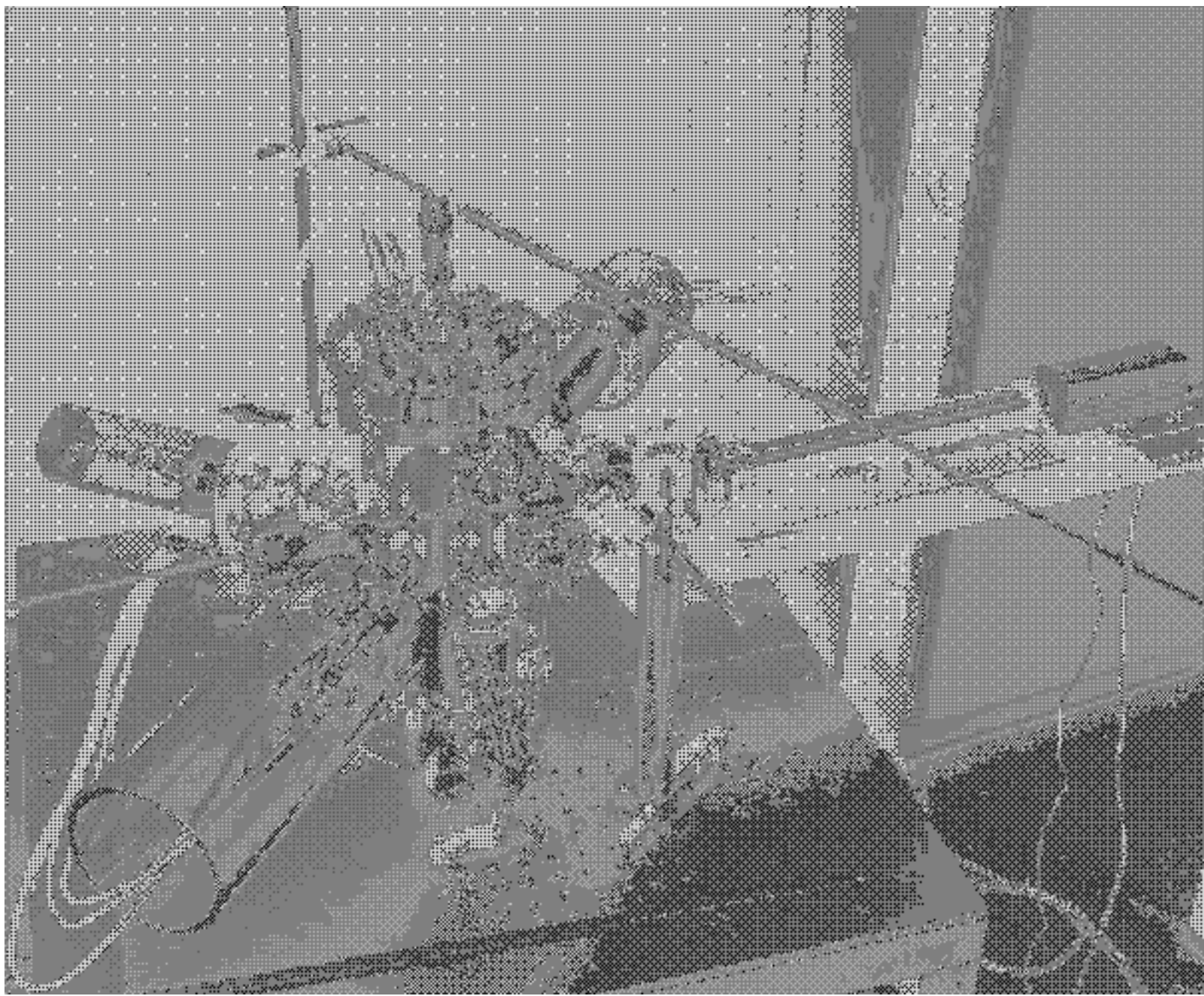
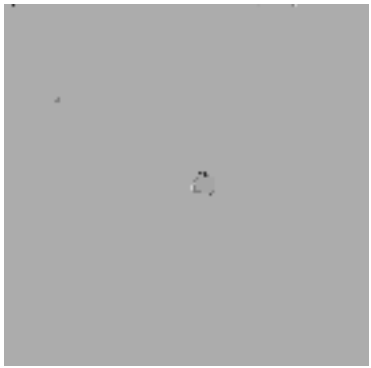
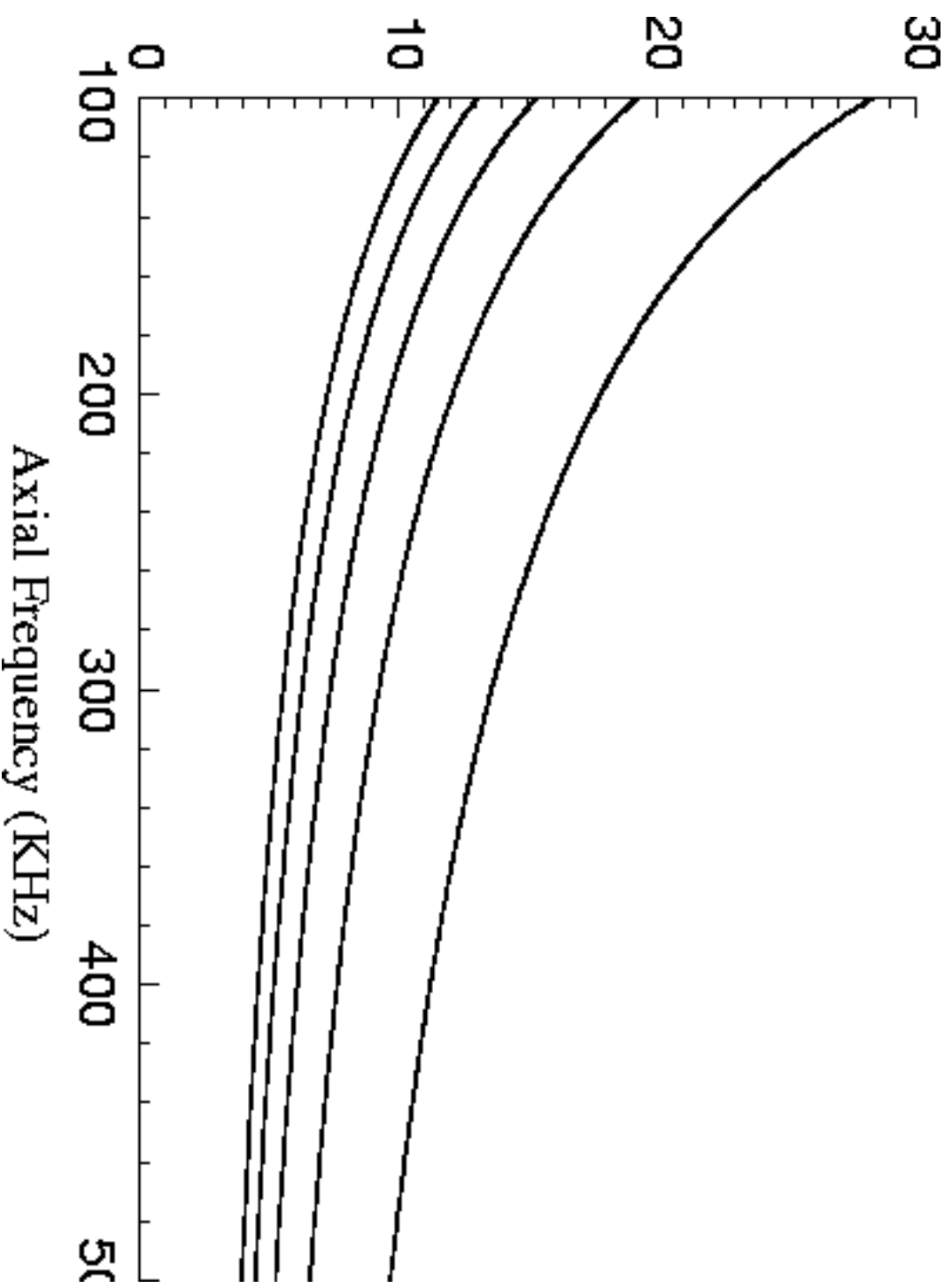
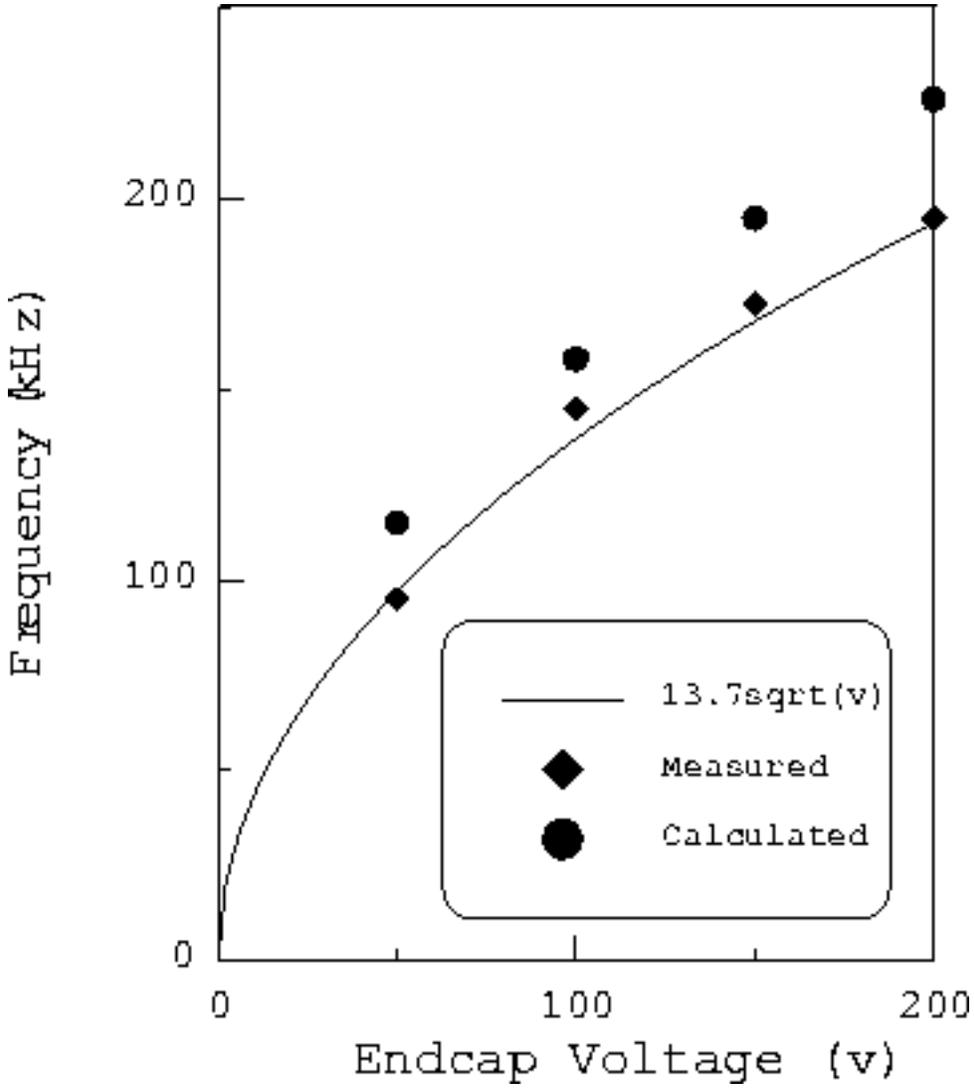
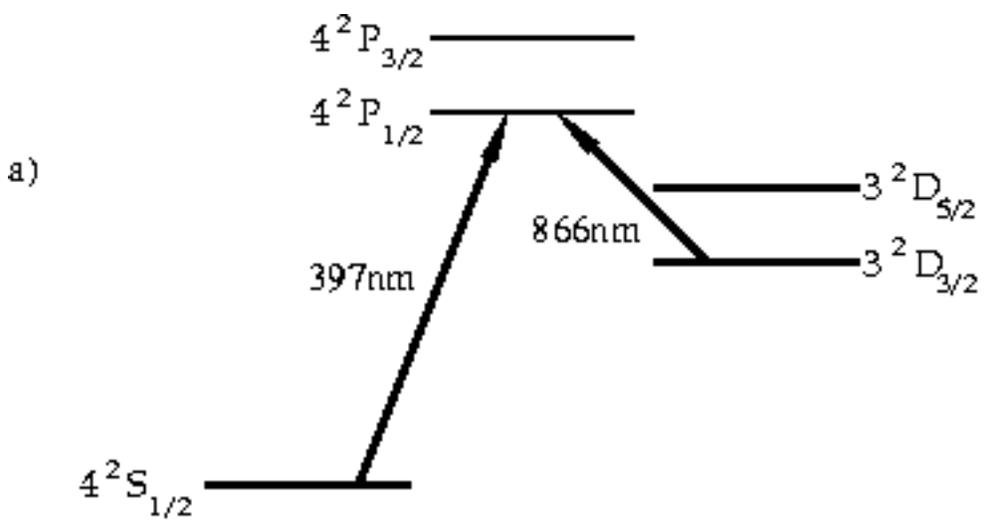
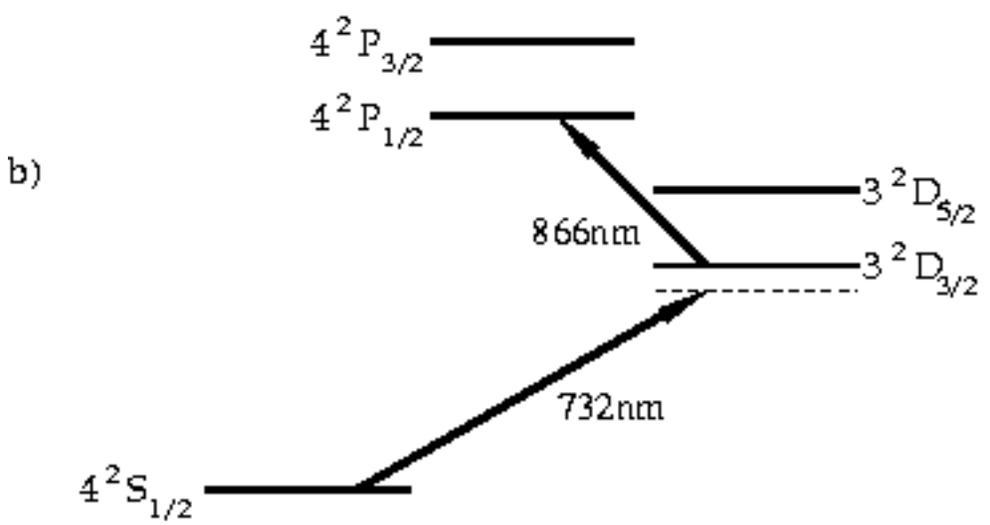
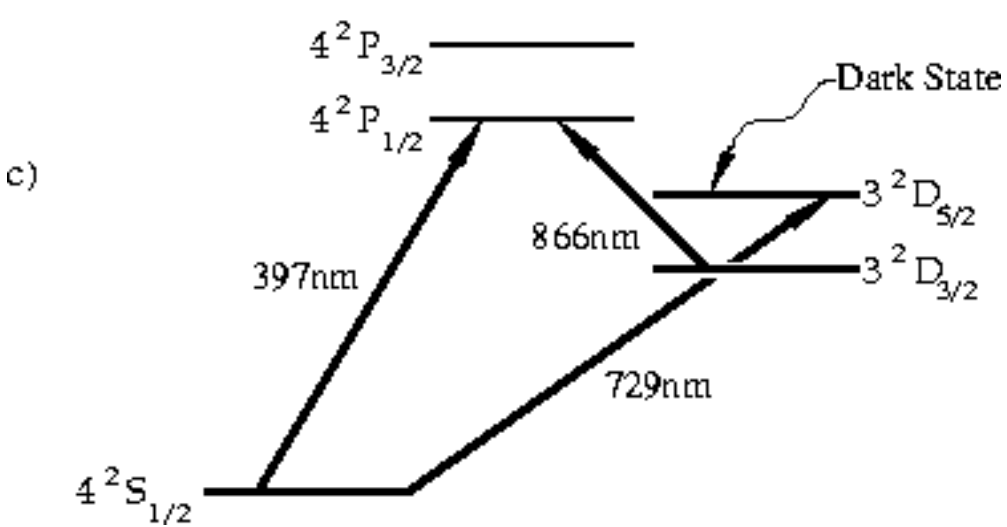
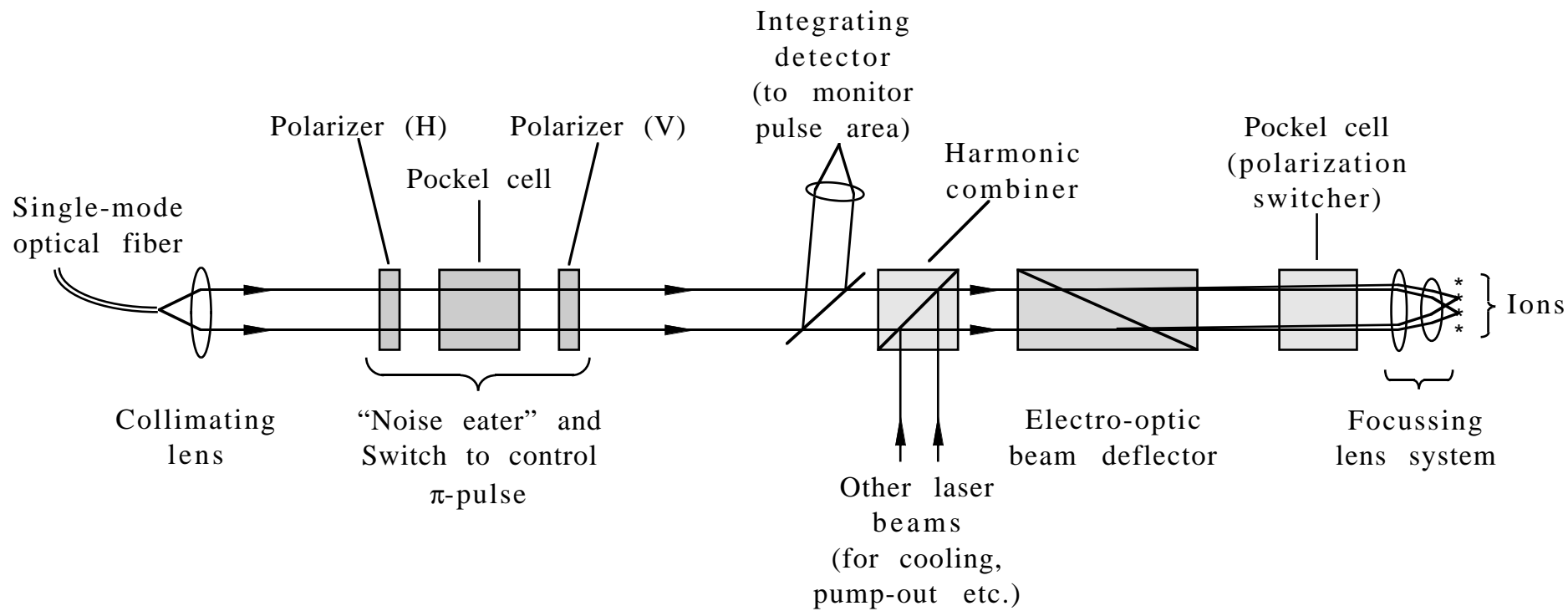
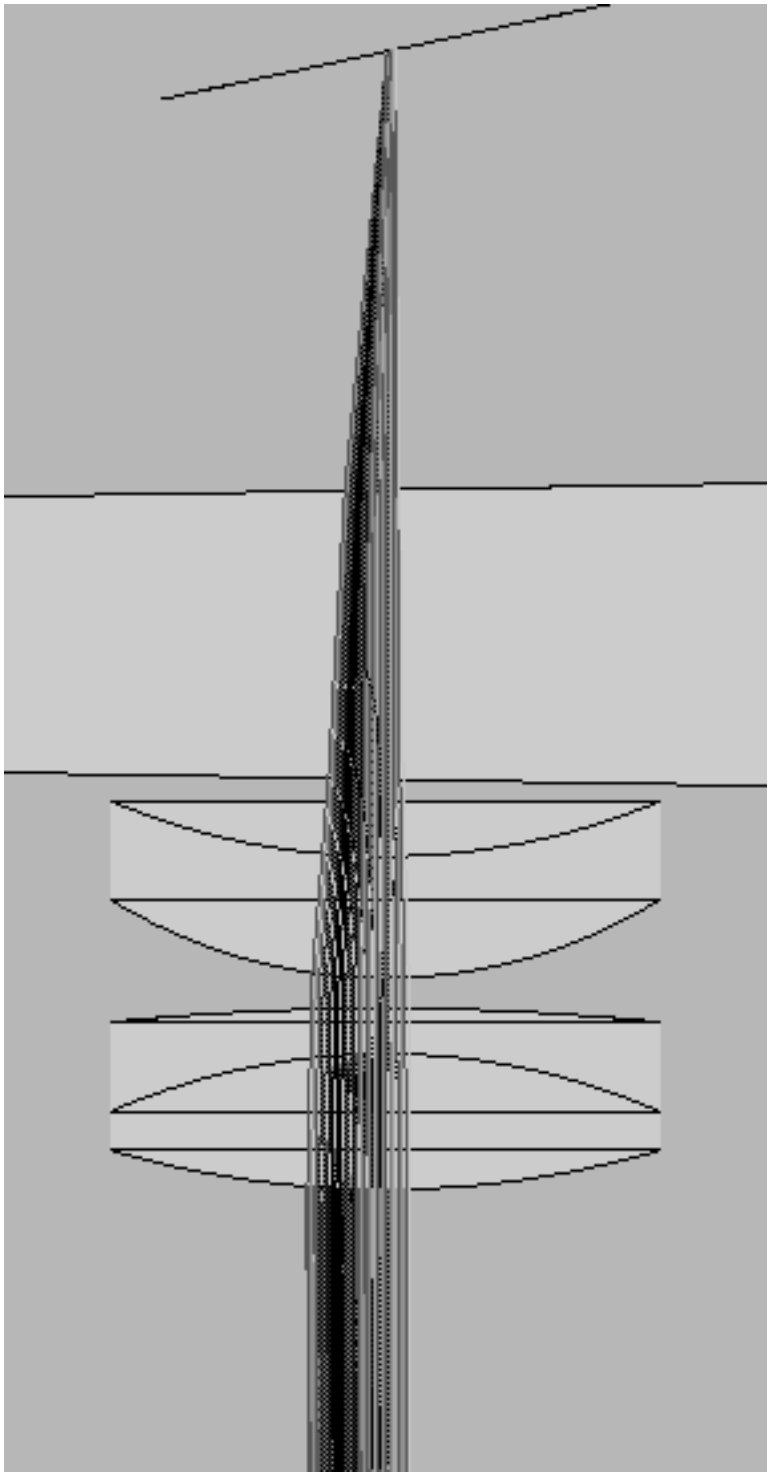
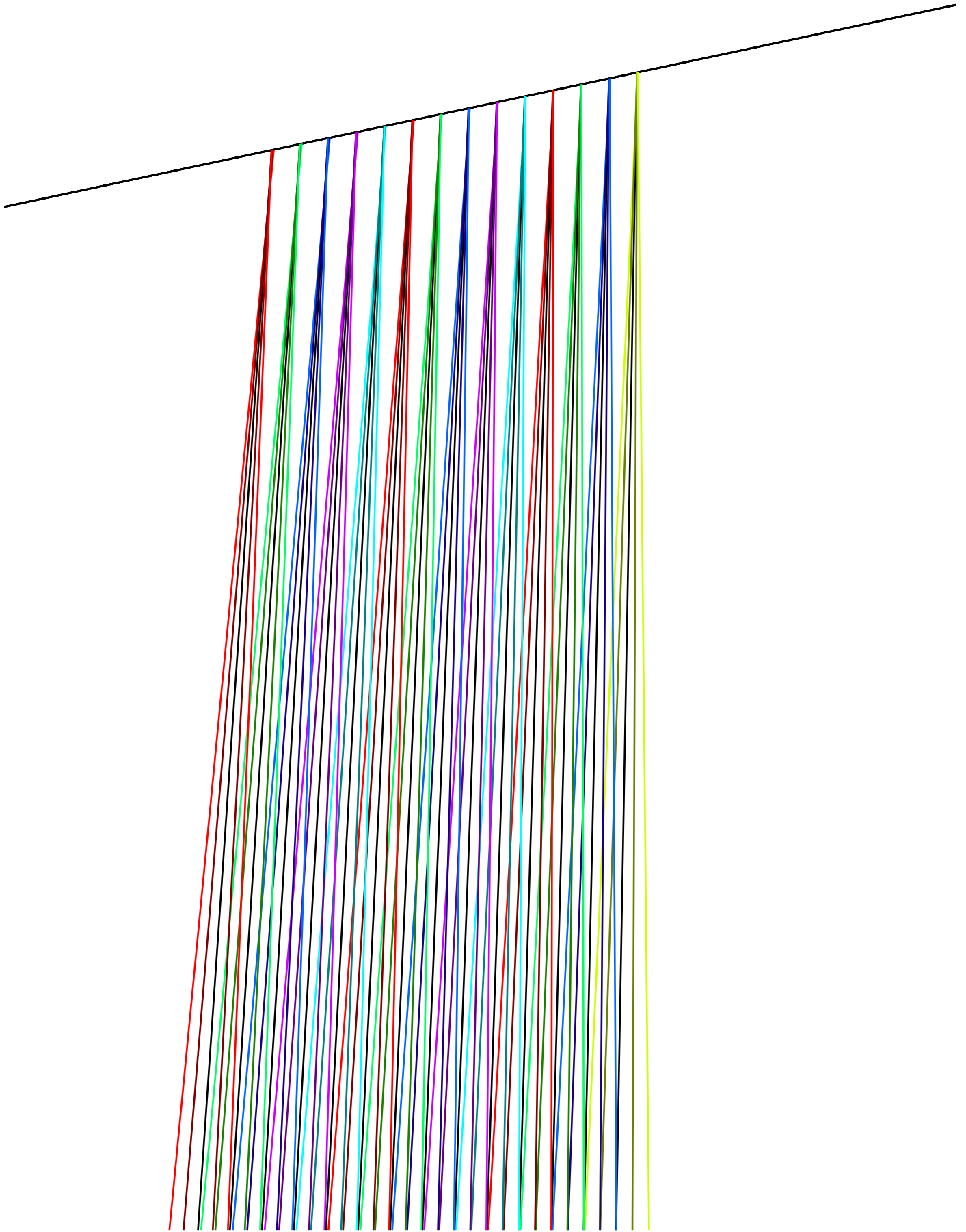
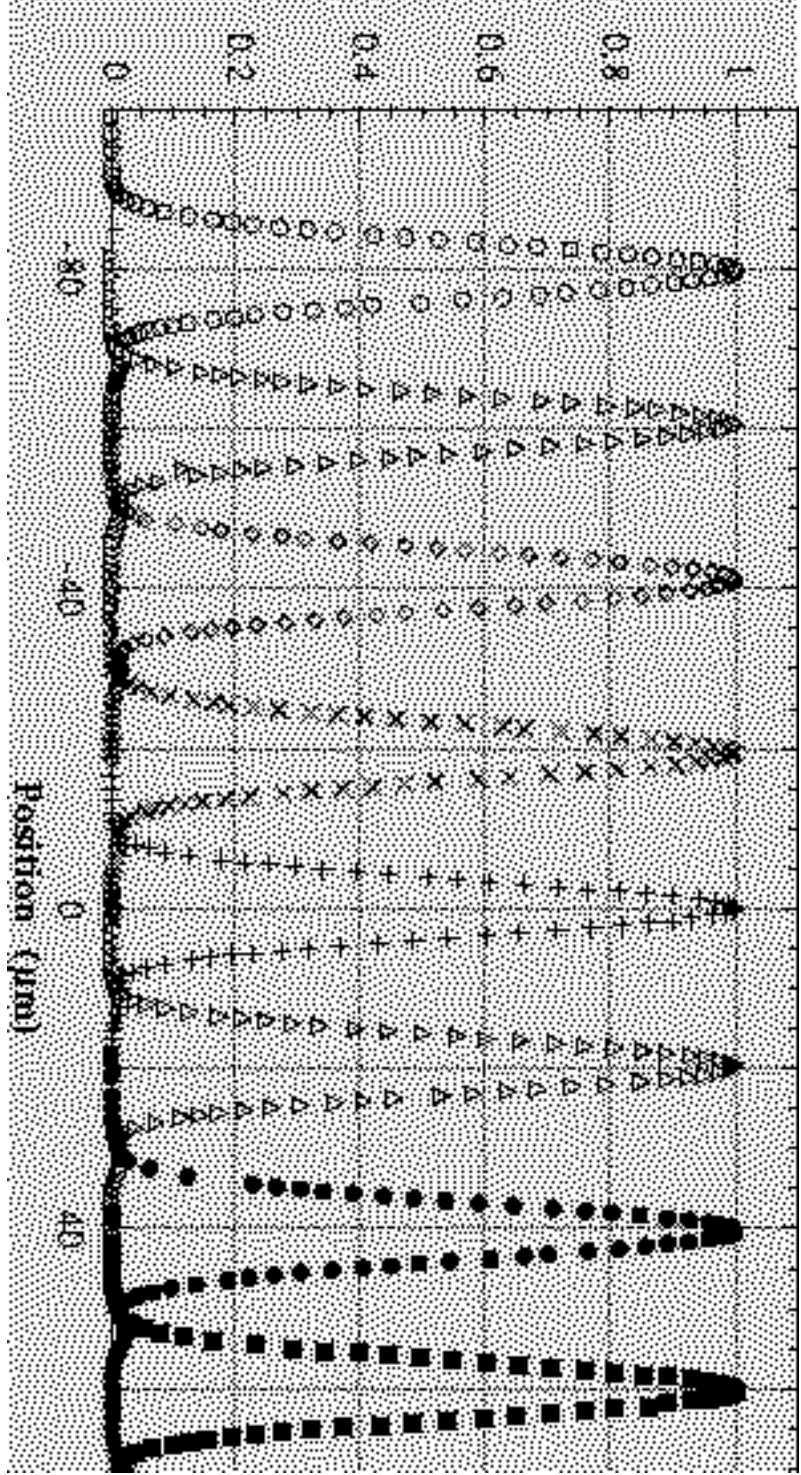
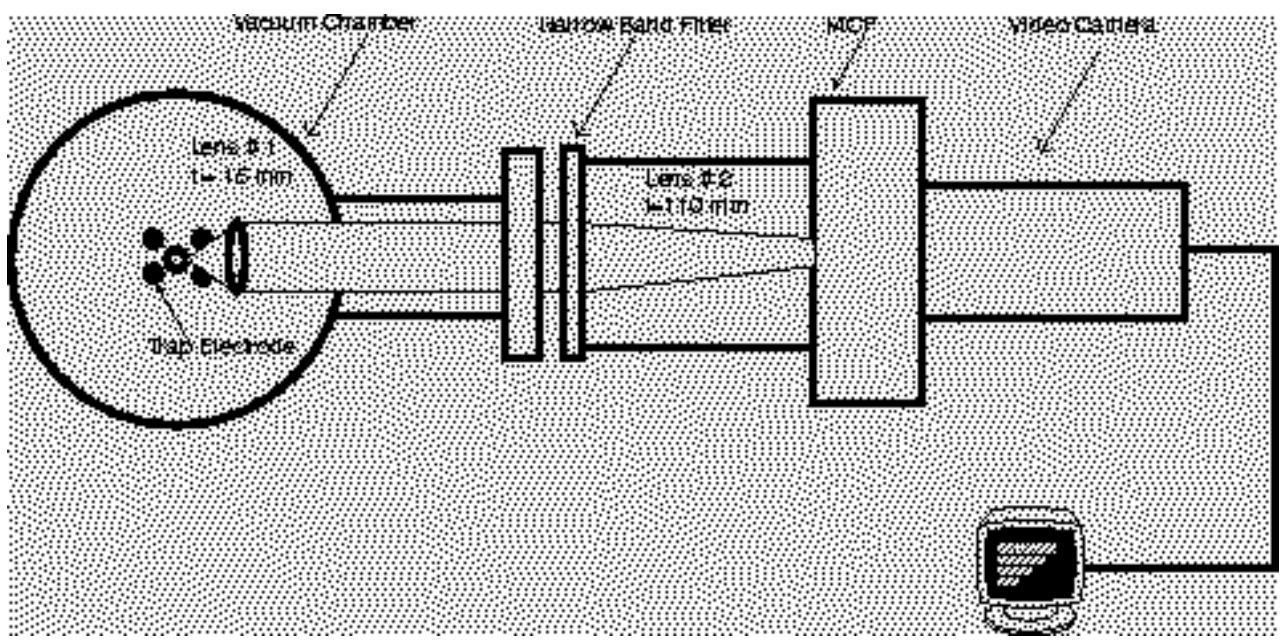
\end{document}